\documentstyle[floats,prd,aps,epsfig,eqsecnum,12pt]{revtex}
\makeatletter
\newbox\tempboxa
\newdimen\captionboxsubcount 
\def\capsize#1{\captionboxsubcount=#1pt}
\newdimen\captionboxsub
\captionboxsub=\hsize \advance\captionboxsub by -\captionboxsubcount
\advance\captionboxsub by -\captionboxsubcount
\long\def\@makecaption#1#2{
 \setbox\@tempboxa\hbox{#1 #2}
 \ifdim \wd\@tempboxa >\captionboxsub 
\rightskip=\captionboxsubcount \leftskip=\captionboxsubcount #1 #2 
\else \hbox to\hsize{\hfil\box\@tempboxa\hfil} 
 \fi}
\makeatother
\capsize{30}

\begin{document}
\bibliographystyle{unsrt}
\begin{titlepage}
\begin{flushright}
\begin{minipage}{5cm}
\begin{flushleft}
\small
\baselineskip = 13pt
SU-4240-711\\
\end{flushleft}
\end{minipage}
\end{flushright}
\begin{center}
\Large\bf

Chiral Lagrangian treatment of $\pi\eta$ scattering 
\end{center}
\vfil
\footnotesep = 12pt
\begin{center}
\large 
Deirdre {\sc Black}
\footnote{Electronic
address: {\tt black@suhep.phy.syr.edu}}
,
\large
\quad\quad Amir H. {\sc Fariborz}
\footnote{Electronic  
address: {\tt amir@suhep.phy.syr.edu}} 
, 
Joseph {\sc Schechter}
\footnote{
Electronic address : {\tt schechte@suhep.phy.syr.edu}}\\
{\it 
\qquad  Department of Physics, Syracuse University, 
Syracuse, NY 13244-1130, USA.} \\
\vskip 0.5cm
\end{center}
\vfill
\begin{center}
\bf
Abstract
\end{center}
\begin{abstract}
We study $\pi\eta$ scattering in a model which starts from the tree
diagrams of a non-linear chiral Lagrangian including appropriate
resonances.  Previously, models of this type were applied to $\pi\pi$ 
and $\pi K$ scattering and were seen to require the existence of light
scalar $\sigma(560)$ and $\kappa(900)$ mesons and to be consistent with the
$f_0(980)$.  The present calculation extends this to include the $a_0(980)$, thereby
completing a possible nonet of light scalars, all ``seen'' in the same
manner.  We note that, at the initial level, the $\pi\eta$ channel is
considerably cleaner than the $\pi\pi$ and $\pi K$ channels for the study
of light scalars.  This is because the large competing effects of vector
meson exchange and ``current-algebra'' contact terms are absent.  The 
simplicity of this channel enables us to demonstrate the 
closeness of our exactly crossing symmetric amplitude to a related exactly 
unitary amplitude.  The calculation is also extended  to higher energies in
order to let us discuss the role played by the $a_0(1450)$ resonance.
\baselineskip = 17pt
\end{abstract}
\begin{flushleft}
\footnotesize
PACS number(s): 13.75.Lb, 11.15.Pg, 11.80.Et, 12.39.Fe 
\end{flushleft}
\vfill
\end{titlepage}
\setcounter{footnote}{0}

\section{Introduction}

In the last few years there has been a revival of interest in the
lowest-lying scalar mesons.  Various authors \cite{Jaffe}-\cite{mechanism} have
debated evidence that their masses lie below $1$ GeV and do not fit the
pattern expected if they were conventional $q \bar q$ states in the quark
model.  This question has also stimulated investigations of meson-meson
scattering, where these particles should be ``seen'' and which anyway is
one of the classical problems of elementary particle physics.  Now it is
accepted that the solution to this problem should look like chiral
perturbation theory very close to the meson-meson scattering threshold.
But when one moves away from threshold into the resonance region, this
essentially polynomial fit to the scattering amplitude requires many higher
order terms and is practically difficult to implement.  An alternative
approach which is the stimulus for the present work, involves retaining
chiral symmetry but using a resonance rather than a polynomial basis for
the scattering amplitude.  This will not be as accurate near threshold but
may be expected to give a simple reasonable picture in the energies up to
the $1$ GeV region.  The $\displaystyle{\frac{1}{N_c}}$ approximation to QCD \cite{1n}
provides a motivation but not a strict proof of this method.  Recently it
has been applied to $\pi \pi$ scattering \cite{Sannino}, $\pi K$ scattering
\cite{Black1} and to ${\eta}^{\prime} \rightarrow \eta \pi \pi$ decays \cite{Fariborz}.  A
light scalar-isoscalar -- $\sigma (560)$ -- and a light scalar-isospinor --
$\kappa (900)$ -- were found to be necessary to fit the $\pi \pi$ and $\pi
K$ scattering data.  The known $f_0(980)$ was also included as a direct
channel resonance in the $\pi \pi$ case and the known $a_0(980)$ was
observed to play an important role in the ${\eta}^{\prime} \rightarrow \eta
\pi \pi$ decay processes.  In the present paper we directly investigate the
$\pi \eta$ scattering process which is expected to feature the $a_0(980)$
as well as the higher scalar isovector $a_0(1450)$.  We are employing the
same method as in the previous treatments in the hope of checking the
validity and improving our understanding of the approach.  

As before, the amplitude will be constructed from a chiral invariant
Lagrangian treated at tree level but ``regularized'' near the direct
channel poles.  The regularization will be regarded as restoring unitarity
in the vicinity of the poles.  The resulting amplitude starts out crossing
symmetric but not exactly unitary.  The burden of the method is to
approximately satisfy unitarity as well as crossing symmetry.  It seems
reasonable to include in the effective Lagrangian all resonances in the
range up to the maximum energy of interest.  We follow this rule up to
about 1 GeV but, for simplicity, only keep the clearly relevant direct
channel pole $a_0(1450)$ above this energy.  

The $\pi \eta$ scattering actually does turn out to contain some
interesting differences from the $\pi \pi$ and $\pi K$ cases.  In those
cases the direct poles $\sigma (560)$ and $\kappa(900)$ had to have a
modified Breit-Wigner shape, with an extra parameter, in order to fit the
experimental data.  This is not unreasonable since these resonances appear
in direct competition with the large ``current-algebra'' contact terms, as
well as strong vector meson exchange contributions, and thus do not
dominate their own channels.  As already seen in the discussion of
${\eta}^{\prime} \rightarrow \eta \pi \pi$ decay \cite{Fariborz}, vector
meson exchanges and ``current algebra'' contact terms do not contribute to
$\pi \eta$ scattering in the elastic region.  Thus it is appropriate to use
the ordinary Breit-Wigner unitarization for the $a_0(980)$. 

In Section II we give our notation for the scattering problem and treat
$\pi \eta$ scattering in the region where the elastic approximation seems
reasonable.  This leads to a description of the $a_0(980)$ which is
consistent with recent experimental data \cite{Teige}.  Section III contains a
discussion of the problem of unitarizing the $I=J=0$ partial wave amplitude
and a comparison of the unitary amplitude with our crossing symmetric one.
In Section IV the $\pi \eta$ channel is discussed in the inelastic region,
up to about $1.6$ GeV, which contains the $a_0(1450)$ resonance.  The model
was simplified by leaving out a number of higher mass resonance
crossed-channel exchanges.   A start on the more involved problem of
including other channels is made in Section V.  Multi-channel unitarity is
checked for the simplified model.  A brief summary and directions for further
work are given in Section VI.  Appendix A gives our chiral Lagrangian and
numerical parameters, while Appendix B shows the elastic $I=J=0$ $\pi \eta$
partial wave amplitude to interested readers.

\section{Elastic $\pi\eta\rightarrow \pi\eta$ scattering amplitude}

We now discuss $\pi \eta$ scattering in a similar way to that previously
employed for $\pi \pi$ and $\pi K$ scattering.  In this section we limit
attention to the energy range (up to roughly 1.2 GeV) where the elastic
approximation seems at least qualitatively reasonable. 

The amplitude, as in the previous treatments, will be obtained from the
tree graphs of the chiral Lagrangian of pseudoscalars, vectors and scalars
given in Appendix A.  This is motivated by the large $N_c$ approximation to
QCD.  As is well known experimentally, the low energy $\pi \eta$ scattering
is dominated by the $a_0(980)$ scalar resonance.  

In detail the $\pi \eta$ scattering turns out to be much simpler than the
$\pi \pi$ and $\pi K$ \cite{Sannino,Black1} cases.  In those examples the leading contributions
near threshold were due to the so-called current algebra contact term (from
the first term of Eq. (\ref{Lag_psv})) and the terms associated with vector meson
exchange.  It is easy to see, using G-parity and isospin conservation, that
there are no vector meson exchanges possible for $\pi \eta \rightarrow \pi
\eta$ scattering at tree level.  Similarly the pseudoscalar contact
contribution (first term of Eq. (\ref{Lag_psv})) which has the same structure as
the vector meson exchange contribution, vanishes identically for $\pi \eta$
scattering.  

\begin{figure}
\centering
\epsfig{file=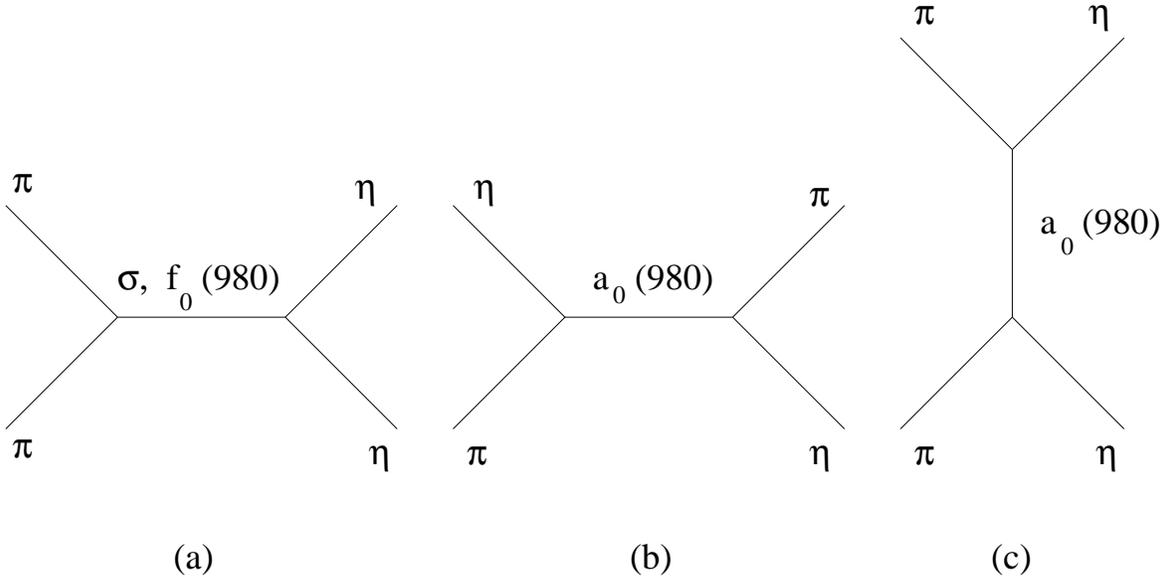, height=3in, angle=0}
\caption
{
Feynman diagrams representing the contributions to the $\pi \eta
\rightarrow \pi \eta$ scattering amplitude of scalar mesons $\sigma(560)$
and $f_0(980)$, (a), $a_0(980)$ in the u-channel, (b) and in the s-channel (c).
}
\label{FD_pe2pe}
\end{figure}

Thus, considering at first only exchanges of particles less than about 1
GeV in mass, restricts us to the scalar mesons.  This process then provides
a very clean channel for studying the scalar meson properties.  Feynman
diagrams, representing the contribution of the scalar mesons to the
scattering amplitude, are shown in Fig. \ref{FD_pe2pe}.  The tree
level invariant amplitude is then simply 

\begin{eqnarray}
A_{\pi \eta} &=&
\sum_{r=\sigma,f_0} \frac{\gamma_{r\pi\pi}\gamma_{r\eta\eta}}{\sqrt 2} \frac{
\left(t-2m_\pi^2\right) \left(t-2m_\eta ^2  \right)}{m_r^2 - t}  
+ \frac {{\gamma_{a\pi\eta}}^2}{4} \frac {{\left(u - m_{\pi}^2 -
m_{\eta}^2 \right)}^2}{m_a^2 - u} \nonumber \\
&+& \frac {{\gamma_{a\pi\eta}}^2}{4} \frac {{\left(s - m_{\pi}^2 -
m_{\eta}^2 \right)}^2}{m_{a_0}^2 - s} 
+2\frac{m_\pi^2}{F_\pi^2}{{\rm cos}^2{\theta_p}}.
\label{APE}
\end{eqnarray}
Here $s$, $t$ and $u$ are the usual Mandelstam variables.  The $\gamma$'s,
scalar $\rightarrow$ pseudoscalar-pseudoscalar coupling constants, were
numerically determined in previous papers \cite{Sannino,Black1,Black2,Fariborz}.
Finally, the last term in Eq. (\ref{APE}) is a small correction which arises from
the pseudoscalar symmetry breaker of Eq. (\ref{PSSB}) and involves the $\eta
-{\eta}^{\prime}$ mixing angle $\theta_p$.  Numerical values of this and other relevant
parameters are listed in Appendix A.  Note that the momentum dependences in
the numerators of the scalar exchange diagrams originate from the chiral
symmetric interaction of Eq. (\ref{Lag_scalars}).

The structure of this amplitude is similar to that for the decay process
${\eta}^{\prime} \rightarrow \eta \pi \pi$.  It was found \cite{Fariborz}
that an appropriate choice of the parameters $C$ and $D$ in
Eq. (\ref{Lag_scalars}) was able to explain the Dalitz plot and overall rate
for the decay process ($A$ and $B$ had previously been found from $\pi \pi$
and $\pi K$ scattering \cite{Black2}).  

An important check of the amplitude is that it obeys the general crossing
symmetry relation:
\begin{equation}
A_{\pi \eta} \left( s,t,u \right) = A_{\pi \eta} \left( u,t,s \right).
\label{crossing}
\end{equation}
While this maps the physical $\pi \eta$ scattering region to an unphysical
one, it is a restriction on the analytic form of $A_{\pi \eta} \left( s,t,u
\right)$.  

Of course, as written, Eq. (\ref{APE}) cannot be meaningfully compared with
experiment for a range of energy beyond 1 GeV because there is a physical
divergence at $s=m_{a_0}^2$.  A usual way to handle this problem is to
``regularize'' the expression by making the substitution:
\begin{equation}
\frac{1}{m_{a_0}^2 - s} \longrightarrow \frac{1}{m_{a_0}^2 - s -
im_{a_0}G^{\prime}_{a_0} \theta \left[ s - {\left(m_{\eta} + m_{\pi}
\right)}^2 \right]},
\label{reg}
\end{equation}
where the $\theta$ function guarantees that there is no imaginary part
below threshold.  A similar substitution for $1/\left( m_{a_0}^2 -u
\right)$ which maintains crossing symmetry (Eq. (\ref{crossing})) gives no
imaginary piece since $u \leq {\left( m_{\eta} - m_{\pi}\right)}^2$ in the
physical scattering region.  The quantity $G_{a_a}^{\prime}$ in the elastic
approximation is interpreted as the decay rate for $a_0(980) \rightarrow
\pi\eta$:
\begin{equation}
\Gamma \left( a_0 \rightarrow \pi \eta \right) = \frac {\left| {\bf
p_{\pi}} \right|}{32 \pi m_{a_0}^2} \gamma_{a\pi\eta}^2 {\left(
m_{a_0}^2 - m_{\pi}^2 - m_{\eta}^2 \right)}^2,
\label{awidth}
\end{equation}
where ${\bf p_{\pi}}$ is the final pion's momentum in the $a_0$ rest frame.
Analysis of the ${\eta}^{\prime} \rightarrow \eta \pi \pi$ process fixed
$\Gamma \left( a_0 \rightarrow \pi \eta \right) \approx 65$ MeV while in
the same model (see end of section IV of \cite{Black2}) it was estimated
that $\Gamma \left( a_0 \rightarrow K\bar K \right) \approx 5$ MeV.  Thus
the elastic approximation for $\pi \eta$ scattering seems not too bad even
slightly beyond the $K \bar K$ threshold.  The effect of this small inelasticity
is taken into account by using $G_{a_0}^{\prime} \approx 70$ MeV rather
than 65 MeV.  

To go further in the analysis it is important to discuss the unitarity
constraint on the scattering amplitude.  This is conveniently done with the
aid of the partial wave projections.  Since we are dealing with $\pi \eta$
scattering the projection will always leave us in the I=1 channel (all of $\pi^{+} \eta$, $\pi^{-} \eta$ and $\pi^{0} \eta$ have the same
amplitude).  Considering a more general case for later use, the desired
angular momentum $l$ partial wave amplitude is given by: 

\begin{equation}
T_{ab;l} (s) = \sqrt {\rho_a(s) \rho_b(s)} \int^1_{-1} d{\rm cos} \theta 
\, P_l (\cos \theta) \, A_{ab} (s,t,u),
\label{projected}
\end{equation} 
where $\theta$ is the center of mass scattering angle and 
\begin{equation}
\rho_a(s) = \frac {q_a(s)}{16\pi \sqrt s},
\label{kinematical}
\end{equation}
with $q_a(s)$ the center of mass momentum for channel $a$, containing particles $a_1$ and $a_2$:
\begin{equation}
q_a^2 = \frac{s^2 + {\left( m_{a_1}^2 - m_{a_2}^2 \right)}^2 - 2s \left(
m_{a_1}^2 + m_{a_2}^2 \right)}{4s}.
\label{cofm}
\end{equation}
$A_{ab} (s,t,u)$ stands for the invariant amplitude for channel a
$\rightarrow$ channel b.  The two-body channels of interest are denoted
$1=\pi \eta$, $2=K \bar K$ and $3= \pi {\eta}^{\prime}$.  For the elastic
region, of course, $A_{11}$ is given by $A_{\pi \eta}$ in Eq. (\ref{APE}).  

The partial wave amplitudes $T_{ab;l}$ are related to the S-matrix elements
by 
\begin{equation}
S_{ab;l} = \delta_{ab} + 2iT_{ab;l},
\label{defineS}
\end{equation}
which satisfy (in the two-particle channel dominance approximation) the
unitarity formula
\begin{equation}
\sum_{b} S_{ab;l}S_{cb;l}^* = \delta_{ac}.
\label{Smatrixunitarity} 
\end{equation}
Since the S-matrix is unitary, $\left|S_{ab;l}\right| \leq 1$ for each of
its entries and so obviously Re$(S_{ab;l}) \leq 1$ and Im$(S_{ab;l}) \leq
1$.  This leads to the very important unitarity bounds on the real and
imaginary parts of the partial wave amplitudes:

\begin{eqnarray}
T_{ab;l} &\equiv& R_{ab;l} + iI_{ab;l}, \nonumber \\
\left| R_{ab;l} \right| &\leq& \frac{1}{2}, \nonumber \\
\left| I_{ab;l} \right| &\leq& \frac{1}{2}\left( 1 + \delta_{ab} \right).
\end{eqnarray}
Usually, if one focuses on the $1 \rightarrow 1$ channel, the standard
parameterization of $S_{11}$ is 
\begin{equation}
S_{11;l} = \eta_l e^{2i\delta_l},
\end{equation}
where $0 < {\eta}_l \leq 1$ is the elasticity parameter and ${\delta}_l$ is
the phase shift, in this case for $\pi \eta \rightarrow \pi \eta$
scattering. 

\begin{figure}
\centering
\epsfig{file=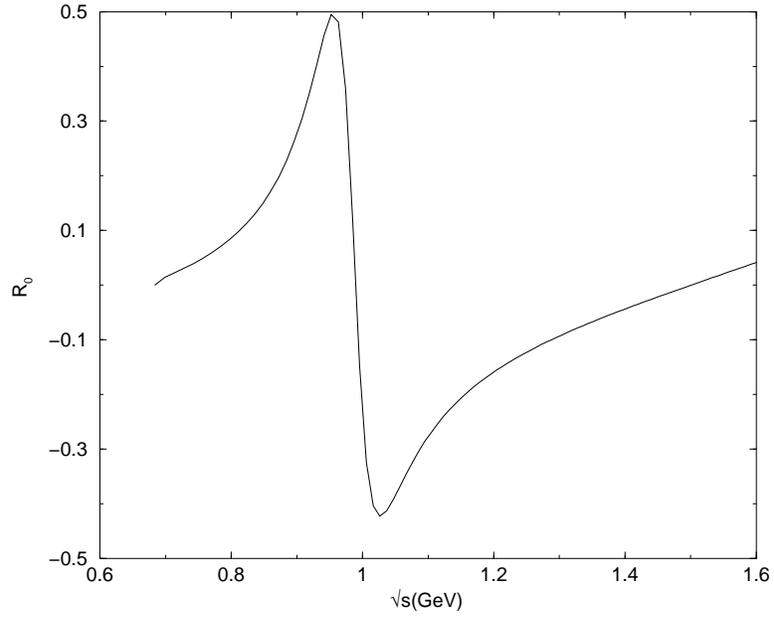, height=4in, angle=270}
\caption
{Plot of the real part of the $l=0$ partial wave projection of
Eq. (\ref{APE}) with the regularization of Eq. (\ref{reg}).
}
\label{pe2pe_bg}
\end{figure}

\begin{figure}
\centering
\epsfig{file=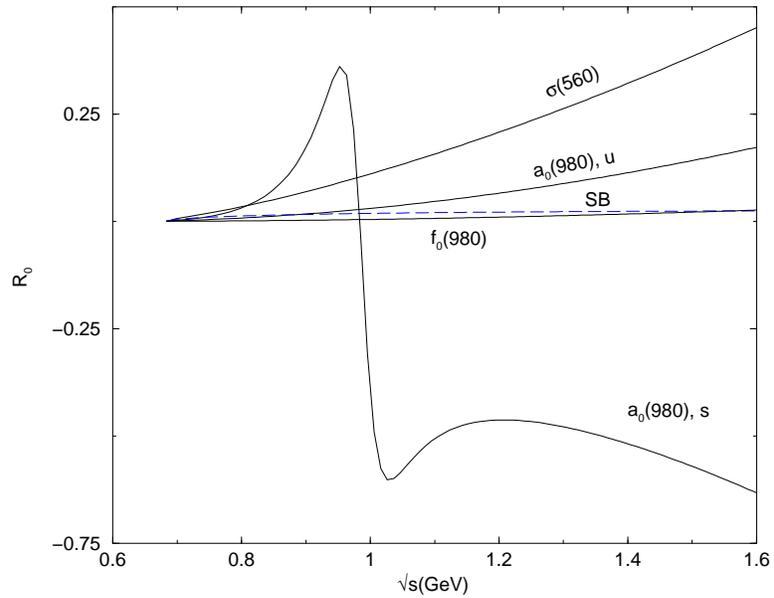, height=4in, angle=270}
\caption
{Individual contributions to $R(s)$ computed as for Fig. \ref{pe2pe_bg}.
}
\label{pe2pe_indv}
\end{figure}

In the present case it is interesting to check the unitarity bound for the
important $l=0$ partial wave amplitude.   For simplicity we drop all
subscripts and denote $T_{11;0} \rightarrow R(s) + i I(s)$.  It is
straightforward to carry out the integration in Eq. (\ref{projected}).  The
results are shown in Appendix B.  A plot of
$R(s)$ up to ${\sqrt s} = 1.6$ GeV, based on taking the $l=0$ partial wave
projection of Eq. (\ref{APE}) with the regularization of Eq. (\ref{reg}) is presented in
Fig. \ref{pe2pe_bg}.  We notice that the unitarity bound $\left| R(s) \right| \leq \displaystyle{\frac{1}{2}}$ is satisfied.  This is not trivial as the plots of the
individual Feynman diagram contributions in Fig. \ref{pe2pe_indv} illustrate.  The
s-channel graph contribution violates unitarity by itself but the $\sigma$ and
other exchanges act to restore the bound.  Nevertheless, the s-channel
graph contribution is clearly the dominant one.  

All in all, the essentially zero parameter prediction shown in
Fig. \ref{pe2pe_bg} seems reasonable.  Of course, satisfying the unitarity bound is a necessary
rather than a sufficient condition for unitarity.  To go further one may
take different points of view.  The simplest is to consider that our
prediction is most accurate and just given for the real part of the
amplitude.  The motivation for this is that the tree graph approximation is
purely real.  Imaginary parts are introduced {\footnote{In the treatments
of $\pi \pi$ and $\pi K$ scattering \cite{Sannino,Black1} the regularizations
for the $\sigma$ and $\kappa$ were introduced as arbitrary parameters which
were varied to make $R(s)$ agree with its experimental shape.  This is very
different from the present $a_0(980)$ case which does {\it{not}} have the
large current algebra and vector meson exchange backgrounds which greatly
complicate the analysis.}} as regularizations like Eq. (\ref{reg}) in the vicinity
of the resonance and may be regarded as (formally small) higher order
effects.  If we then choose to regard only $R(s)$ as predicted, we can
satisfy unitarity by solving the unitarity formula $\left| S \right| =
\left| \eta \right|$, which reads explicitly
\begin{equation}
R^2(s) + {\left[ I(s) - \frac{1}{2} \right]}^2 =
{\left[\frac{\eta(s)}{2}\right]}^2,
\label{unitaritycircle}
\end{equation}
for $I(s)$.  The elasticity factor $\eta(s)$ may be taken to be
approximately one.  This procedure does not exactly solve the problem of
finding an amplitude which satisfies both crossing symmetry and unitarity.
While $R(s)$ coincides with the real part of the $l=0$ projection of the
crossing symmetric Eq. (\ref{APE}), $I(s)$ obtained from
Eq. (\ref{unitaritycircle}) above does not coincide with the imaginary part
of this projection.  We note that the problem of constructing an invariant
amplitude satisfying both crossing symmetry (Eq. (\ref{crossing})) and
unitarity (Eq. (\ref{unitaritycircle})) for all its partial wave projections is
an ancient and difficult one.

\section{More on Unitarity}

\begin{figure}
\centering
\epsfig{file=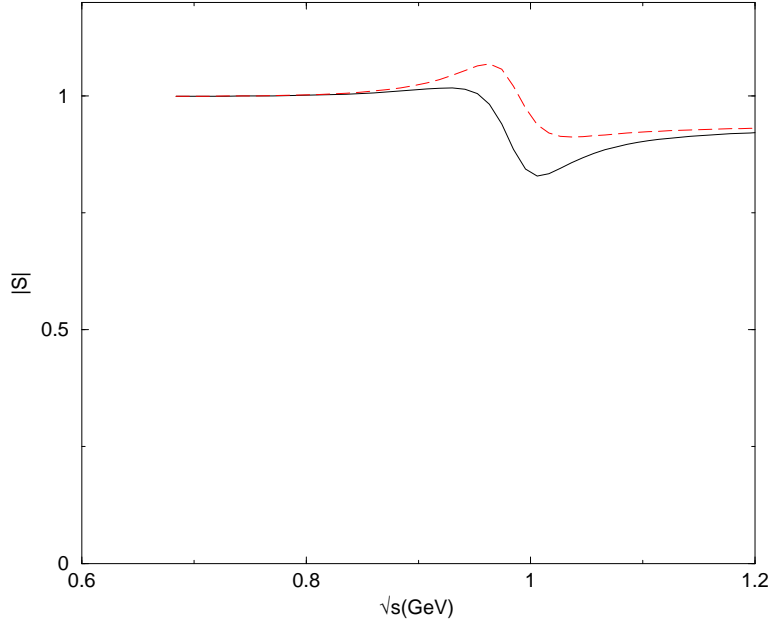, height=4in, angle=270}
\caption
{Unitarity of crossing symmetric S-matrix for $\pi \eta \rightarrow
\pi \eta$ scattering. The dashed line corresponds to the elastic
assumption, $G_{a_0}^{\prime}= 65$ MeV in Eq. (\ref{reg}), while the solid
line takes some inelasticity into account by setting $G_{a_0}^{\prime}=70$
MeV as explained in the text. 
}
\label{ModS_pe2pe}
\end{figure}

In the preceding section we started from the crossing symmetric invariant
amplitude Eq. (\ref{APE}) and regularized it according to the crossing
symmetric prescription Eq. (\ref{reg}).  There was no guarantee that its
$l=0$ partial wave projection would be unitary or even satisfy the unitarity
bounds.  Fortunately the real part $R$ did satisfy the unitary bound and we
could {\it{choose}} an {\it{imaginary}} part $I$ according to
Eq. (\ref{unitaritycircle}) such that partial wave unitarity in the $l=0$
channel was satisfied.  This was at the expense of crossing symmetry for
the imaginary part $I$.  To see by how much the original amplitude
Eq. (\ref{APE}) with the regularization Eq. (\ref{reg}) and its own imaginary part
(i.e. the pure crossing symmetric case) violates unitarity, we have plotted
$|S|$ in Fig. \ref{ModS_pe2pe}.  It is seen that the unitarity violation in
the elastic case (dashed line) is not very
severe;  this is due to the fact that the $a_0(980)$ resonance, which is
approximately of Breit-Wigner shape, dominates.  

If $|R(s)|$ had turned out to be greater than $\displaystyle{\frac{1}{2}}$ at some values of
$s$, we would have of course been unable to impose unitarity with the above
method.  It seems interesting to discuss a more or less conventional type
of unitarization procedure which can be used in such a case.  This operates
at the level of the $l=0$ partial wave amplitude.  We decompose the S-matrix
into a piece associated with the $a_0$ resonance pole, $S_{pole}$, and a
piece associated with the remaining ``background'' terms ($t$ and $u$-channel exchanges and contact
term), $S_B$.  The total S-matrix is written as the product
\begin{eqnarray}
S\left( s \right) &=& S_B\left( s \right) S_{pole} \left( s \right)
\nonumber \\
&=& e^{2i \delta_B\left( s \right)} \frac {m_{a_0}^2 - s + im_{a_0}\Gamma
\left( s \right) }{m_{a_0}^2 - s - im_{a_0}\Gamma\left( s \right) }.
\label{prod}
\end{eqnarray}
In this form unitarity, $|S|=1$, is obvious.  Now, using $S=1+2iT$
yields for the partial wave amplitude,
\begin{eqnarray}
T\left(s \right) &=& T_B \left( s \right) + T_{pole} \left( s \right)
\nonumber \\
&=& e^{i\delta_B\left( s \right)} {\rm{sin}}\delta_B\left( s \right) +
e^{2i\delta_B\left( s \right)} \frac {m_{a_0} \Gamma \left( s
\right)}{m_{a_0}^2 - s - im_{a_0}\Gamma \left(s \right)}.
\label{prod2}
\end{eqnarray}

For orientation, note that when the background phase $\delta_B$ vanishes
and $\Gamma \left( s \right)$ is constant, this reduces to a pure
Breit-Wigner form which is known to be unitary.  If $\delta_B = \overline{
{\delta_B}}$ and $\Gamma = \overline{\Gamma}$ are taken as constants we get
the conventional local form \cite{Taylor} for a narrow resonance in a
constant background:
\begin{equation}
T_{pole} \approx e^{2i\overline{\delta_B}}\frac{m_{a_0} \bar \Gamma}{m_{a_0}^2 -
s - im_{a_0} \bar \Gamma}.
\label{prod3}
\end{equation}

As an example of an application of this last formula, we remark that the
presence of a light $\sigma$ in $\pi \pi$ scattering produces
\cite{Sannino} a background phase $\overline{\delta_B} \approx \displaystyle{\frac {\pi}{2}}$
near the $f_0(980)$.  This results in an overall minus sign which converts
the resonance peak into a dip.  This is the same mechanism -- the
Ramsauer-Townsend effect -- which was observed in atomic scattering
\cite{Taylor} a long time ago.  

Returning to the present case of elastic $\pi \eta$ scattering in the $l=0$
partial wave, we should, to get an exactly unitary amplitude, identify in Eq. (\ref{prod2})
\begin{equation}
\Gamma \left( s \right) = \frac {q}{32 \pi m_{a} \sqrt {s}} \gamma_{a \pi\eta}^2 {\left(
s - m_{\pi}^2 - m_{\eta}^2 \right)}^2,
\label{energydepreg}
\end{equation}
where $q$ is the center of mass momentum.
Furthermore, noting that $T_B(s)$ as obtained from the partial wave
projection of the appropriate terms in Eq. (\ref{APE}) is purely real, we
identify $\delta_B\left( s \right)$ from:  
\begin{equation}
\frac{1}{2} {\rm{sin}} \left[ 2 \delta_B \left( s \right)\right] = R_B
\left( s \right).
\label{backR}
\end{equation}
In order that $T_B$ be unitary we must then {\it{manufacture}} an imaginary
part from (\ref{backR}) as 
\begin{equation}
I_B(s) = {\rm {sin}}^2 \left[ \delta_B \left( s \right) \right].
\label{backI}
\end{equation}
The amplitude so constructed will satisfy $SS^* = 1$ exactly but is,
since among other things we have added $I_B\left( s\right)$ by hand,
expected to violate crossing symmetry. 

\begin{figure}
\centering
\epsfig{file=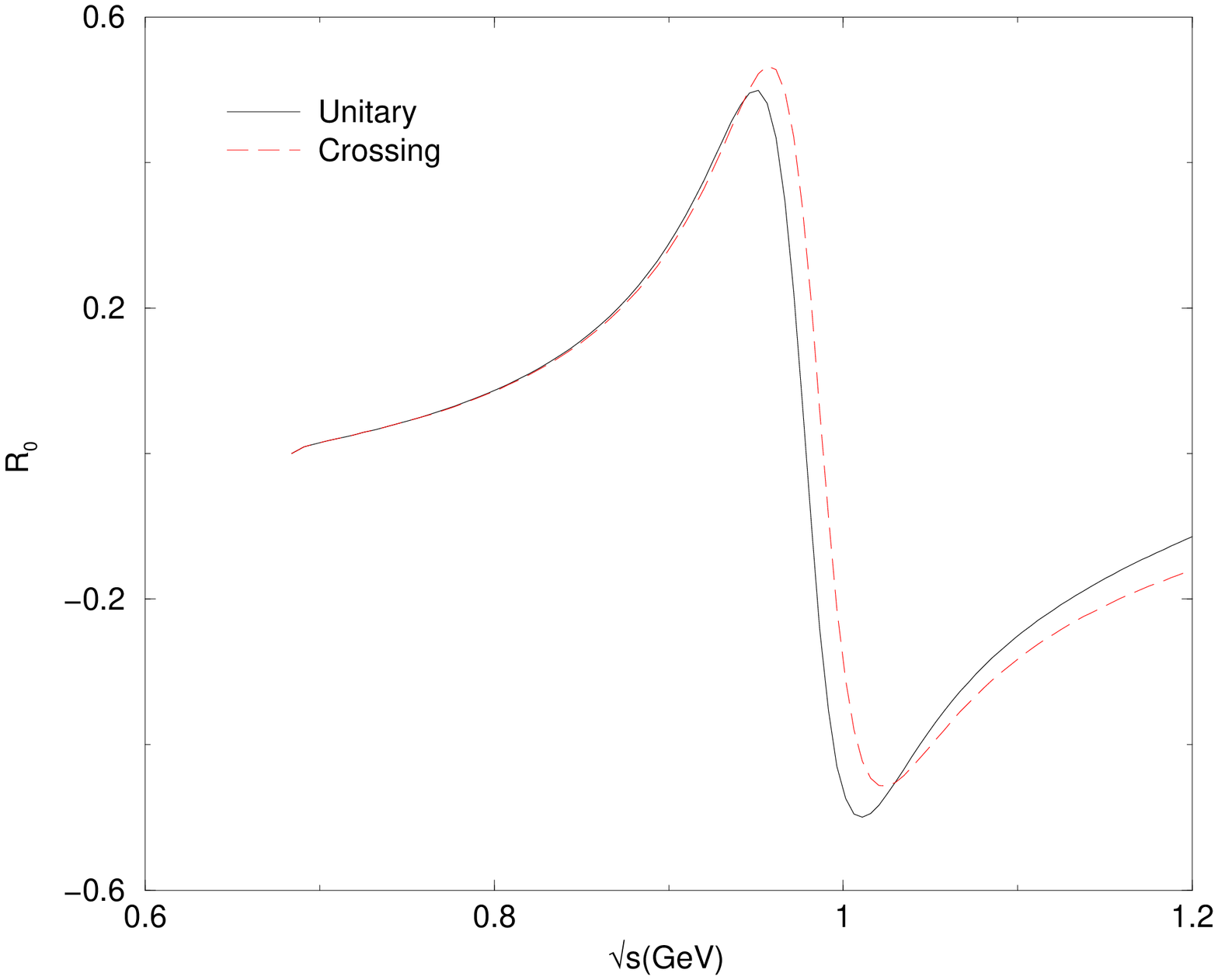, height=5in, angle=270}
\epsfig{file=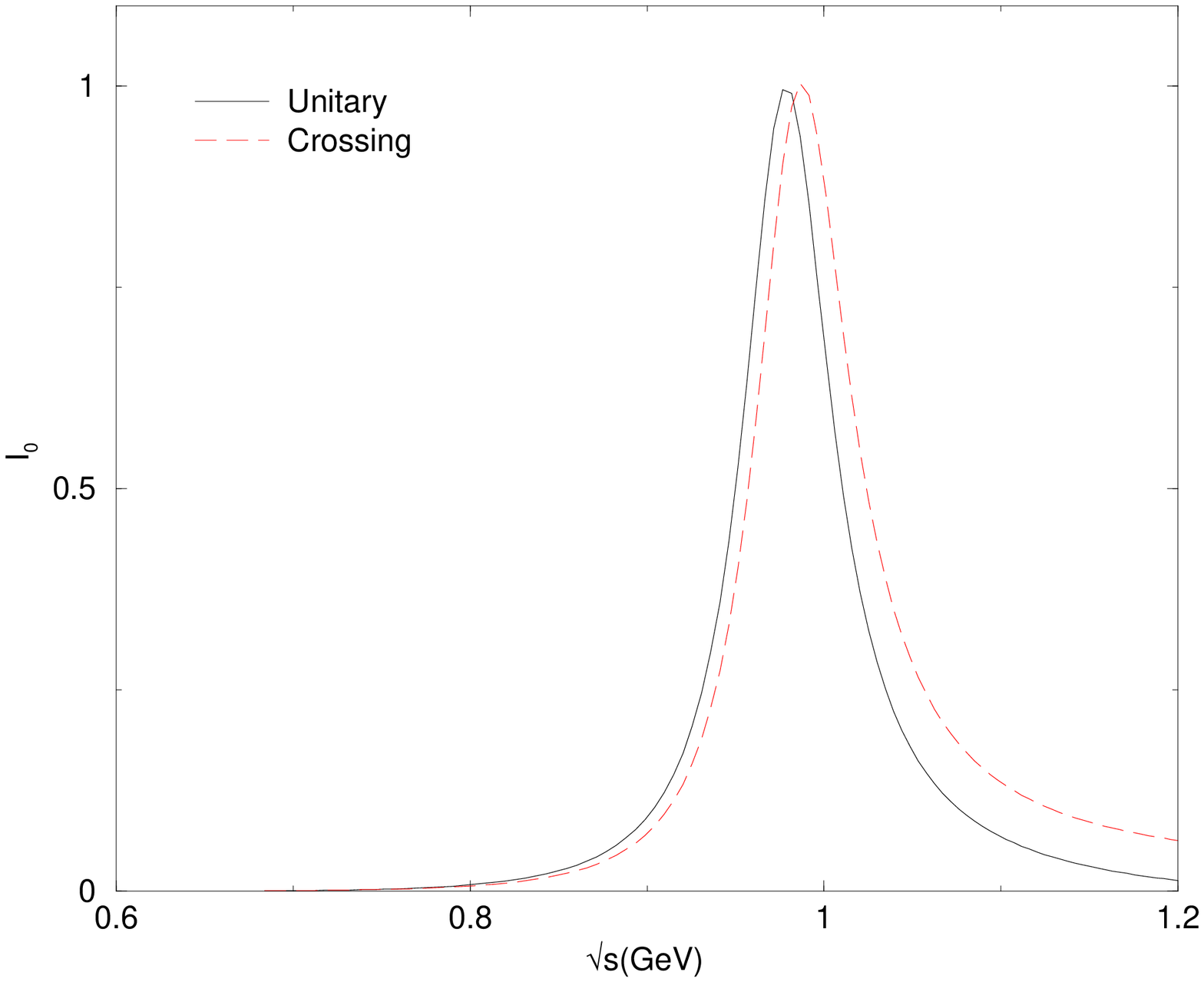, height=5in, angle=270}
\caption
{
Comparison of the real and imaginary parts of the exactly crossing
symmetric amplitude obtained in section II with the exactly unitary
amplitude obtained in section III.
}
\label{pe2pe12_UX}
\end{figure}

To summarize, we compare in Fig. \ref{pe2pe12_UX} the real and imaginary
parts of the $l=0$ projection of the crossing symmetric amplitude obtained in section
II with the exactly unitary amplitude just discussed.  It is encouraging
for the method that the crossing symmetric but not exactly unitary
amplitude is close to the unitary but not exactly crossing symmetric amplitude.  Reasonably, the difference between the two curves gives a measure
of the systematic uncertainties in the present method.  Of course, there is
no guarantee that the true solution lies between the two curves.  

\section{$\pi \eta$ scattering in the inelastic region}

Here we give a start on extending the previous analysis to $\pi \eta$
scattering in the 1.2-1.6 GeV region, where the effects of inelastic
channels, namely $\pi \eta \rightarrow K \bar K$ and $\pi \eta \rightarrow
\pi {\eta}^{\prime}$, are expected to become important.  For a full
description we should use a coupled three-channel approach.  According to
our initially stated picture we should also include all of the resonance
multiplets up to about 1.6 GeV as exchange particles.  This amounts to a
fairly large number of resonances {\footnote{In our treatment we shall
first neglect exchanges of the spin two nonet which includes $f_2(1270)$,
$a_2(1320)$, $K_2^*(1430)$ and $f_2^{\prime}(1525)$.  We shall also neglect
the isoscalar spin zero resonances $f_0(1370)$, $f_0(1500)$ and
$f_J(1710)$, whose properties are not yet definitively known.  Finally in
the $\pi \eta \rightarrow K \bar K$ reaction we shall neglect the
$K^*(1410)$ and $K_2^*(1430)$ resonances.}} (including some whose masses
and decay properties are not yet firmly established) and suggests that such
an ambitious program be postponed.  In the present paper we shall survey
the situation in analogy to an earlier treatment of $\pi K$ scattering in
\cite{Black2} by concentrating on the $\pi \eta \rightarrow \pi \eta$
channel.  In the 1.2-1.6 GeV region the $a_0(1450)$ is the only scalar
resonance in the direct channel.  We shall also compute the $\pi \eta \rightarrow K
\bar K$ and $\pi \eta \rightarrow \pi {\eta}^{\prime}$ amplitudes in our
model, including the effects of the $a_0(1450)$.  For these off-diagonal
transitions we shall be mainly content to check unitarity.  

Our approximation for the $\pi \eta \rightarrow \pi \eta$ amplitude up to
about 1.6 GeV then consists of the sum of the amplitude given in section
II and a piece associated with the $a_0(1450)$ resonance.  The piece in
section II is the $l=0$ partial wave projection of the invariant amplitude in
Eq. (\ref{APE}), regularized according to Eq. (\ref{reg}).  This gives an exactly
crossing symmetric partial wave amplitude which we also saw in section III
to be not very different from an exactly unitary partial wave amplitude.  The regularized invariant amplitude
including the $a_0(1450)$ is taken to be:
\begin{equation}
A_{\pi \eta} \left( s,t,u \right) \approx ... + \frac
{\gamma_{a^{\prime}\pi\eta}^2}{4} \frac {{\left( u - m_\pi^2 - m_\eta^2
\right)}^2} {m_{a^{\prime}}^2 -u} + e^{2i\overline{\delta_{a^{\prime}}}}  \frac
{\gamma_{a^{\prime}\pi\eta}^2}{4} \frac {{\left( s - m_\pi^2 - m_\eta^2
\right)}^2} {m_{a^{\prime}}^2 -s - im_{a^{\prime}} \Gamma_{tot} \left(
a^{\prime} \right)},
\label{1450}
\end{equation}
where $a^{\prime}$ denotes the $a_0(1450)$ and the ellipsis stands for the
other terms just mentioned.  The coupling constants are related to the
widths by an obvious generalization of Eq. (\ref{awidth}).  We have illustrated how to
regularize the s-channel pole term as in (\ref{prod3}) so as to maintain
unitarity near where this term would have blown up.  The phase $\overline{
\delta_{a^{\prime}}}$ is evaluated from 
\begin{equation}
\frac{1}{2} {\rm {sin}} \left[ 2 \overline {\delta_{a^{\prime}}} \right] = R
\left( s= m_{a^{\prime}}^2 \right),
\end{equation}
where $R(s)$ is the real part of the partial wave amplitude due to the
background supplied by the regularized Eq. (\ref{APE}).  We could multiply the u-channel term also by 
$e^{2i\overline{\delta_{a^{\prime}}}}$ to force crossing symmetry.  Actually, we
see from Fig. {2} that  $R\left( s= m_{a^{\prime}}^2 \right)$ is zero to
good accuracy so this correction is not needed in the present case.  Fig. 3
shows that this simplification arises because of cooperation between the
pole and exchange terms.  With $\overline {\delta_{a^{\prime}}}=0$,
Eq. (\ref{1450}) is manifestly crossing symmetric (see the remark after
Eq. (\ref{reg})).  

Note that Eq. (\ref{1450}) explicitly takes some account of inelasticity
since $\Gamma _{tot} \left( a^{\prime} \right) = 265 \pm 13$ MeV \cite{PDG}
is not just gotten from the $a_0(1450) \rightarrow \pi \eta$ width as in
the generalization of Eq. (\ref{awidth}) but will be computed as 
\begin{equation}
\Gamma_{tot} \left( a^{\prime} \right) \approx \Gamma_{\pi \eta} \left(
a^{\prime} \right) + \Gamma_{K \bar K} \left(
a^{\prime} \right) + \Gamma_{\pi {\eta}^{\prime}} \left(
a^{\prime} \right).
\end{equation}
This corresponds to the assumption that the $\pi \eta$, $K \bar K$ and $\pi
{\eta}^{\prime}$ decay modes listed in \cite{PDG} saturate the $a_0(1450)$
decay, although that assumption is not yet confirmed experimentally.  If the
experimental ratios \cite{PDG}
\begin{equation}
\frac{\Gamma_{K\bar K}}{\Gamma_{\pi\eta}} = 0.88 \pm 0.23
\end{equation}
and 
\begin{equation}
\frac{\Gamma_{\pi {\eta}^{\prime}}}{\Gamma_{\pi\eta}} = 0.35 \pm 0.16
\end{equation}
are taken together with this assumption we deduce:
\begin{eqnarray}
\Gamma \left[ a_0(1450) \rightarrow \pi \eta \right] &\approx& 119 \pm 26
\, {\rm {MeV}}, \nonumber \\
\Gamma \left[ a_0(1450) \rightarrow K \bar K \right] &\approx& 105 \pm 36 \,
{\rm {MeV}}, \nonumber \\
\Gamma \left[ a_0(1450) \rightarrow \pi {\eta}^{\prime} \right] &\approx&
42 \pm 23 \,
{\rm {MeV}}.
\label{PDE}
\end{eqnarray}
Actually these three partial widths should be related to each other by
flavor SU(3) invariance.  A best fit, on this assumption, yields the
slightly different central values:  
\begin{eqnarray}
\Gamma^{\rm{SU}(3)} \left[ a_0(1450) \rightarrow \pi \eta \right] &\approx& 155
\, {\rm {MeV}}, \nonumber \\
\Gamma^{\rm{SU}(3)} \left[ a_0(1450) \rightarrow K \bar K \right] &\approx& 86
\, {\rm {MeV}}, \nonumber \\
\Gamma^{\rm{SU}(3)} \left[ a_0(1450) \rightarrow \pi {\eta}^{\prime} \right]
&\approx& 24 \, {\rm {MeV}}.
\label{PDT}
\end{eqnarray}
A more detailed discussion of the $a_0(1450)$ decay modes is given in
\cite{mechanism}.

\begin{figure}
\centering
\epsfig{file=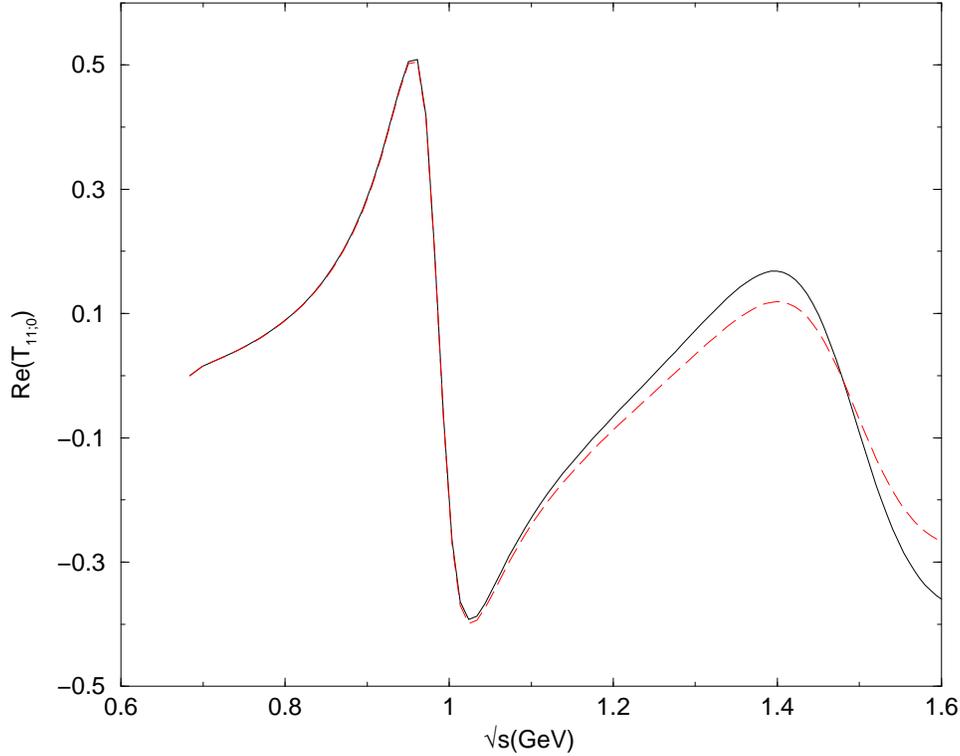, height=5in, angle=270}
\caption
{Our prediction for the real part of the $\pi \eta \rightarrow \pi \eta$
scattering amplitude up to 1.6GeV.  The case where the decay properties of
the $a_0(1450)$ are taken from Eq. (\ref{PDT}) is represented by the solid
line, and may be compared with the case using the estimate Eq. (\ref{PDE})
shown by the dashed line.
}
\label{pe2pe_Total}
\end{figure}

\begin{figure}
\centering
\epsfig{file=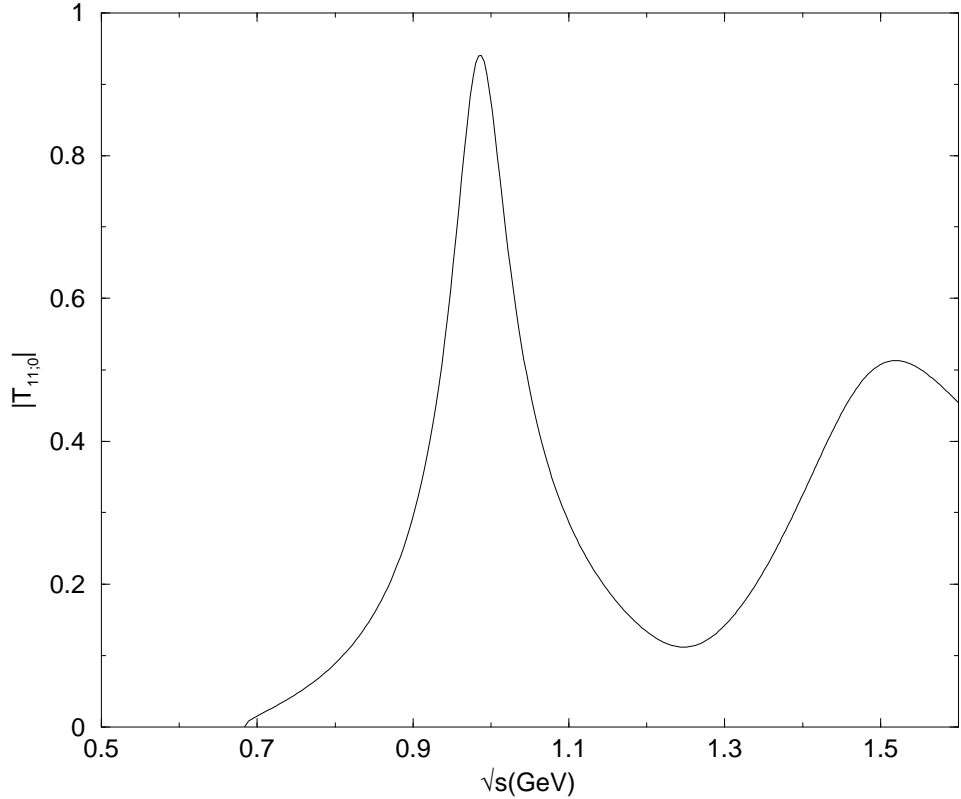, height=5in, angle=270}
\caption
{Modulus of the $\pi \eta \rightarrow \pi \eta$ scattering amplitude.}
\label{ModT11}
\end{figure}

The real part of the $l=0$ partial wave amplitude is plotted up to 1.6 GeV
using both Eq. (\ref{PDE}) and Eq. (\ref{PDT}) in Fig. \ref{pe2pe_Total}.  We see that these two
cases are relatively close{\footnote{In subsequent plots we shall
continue the same convention where the solid line represents the
Eq. (\ref{PDT}) determination and the dashed line represents the Eq. (\ref{PDE})
determination.}}.  Our prediction for the real part of the scattering
amplitude naturally remains within the allowed range of -0.5 to 0.5 (except
for a negligible deviation near the location of the $a_0(980)$).  We have
also plotted in Fig. \ref{ModT11} $\left| T_{11} \right| = \left| T \left( \pi \eta \rightarrow
\pi \eta \right) \right| = \left| R_{11} + i I_{11} \right|$.  This also is
seen to satisfy the unitarity bound $\left| T_{11} \right| \leq 1$.  It
thus seems that the $\pi \eta \rightarrow \pi \eta$ scattering channel is
remarkably simple in the approximation where the $a_0(1450)$ describes the
inelastic region around 1.5 GeV.  The partial wave $l=0$ amplitude is
obtained as a projection of an exactly crossing symmetric invariant
amplitude and the unitarity bounds are satisfied. 

\section{Off-diagonal channels}

\begin{figure}
\centering
\epsfig{file=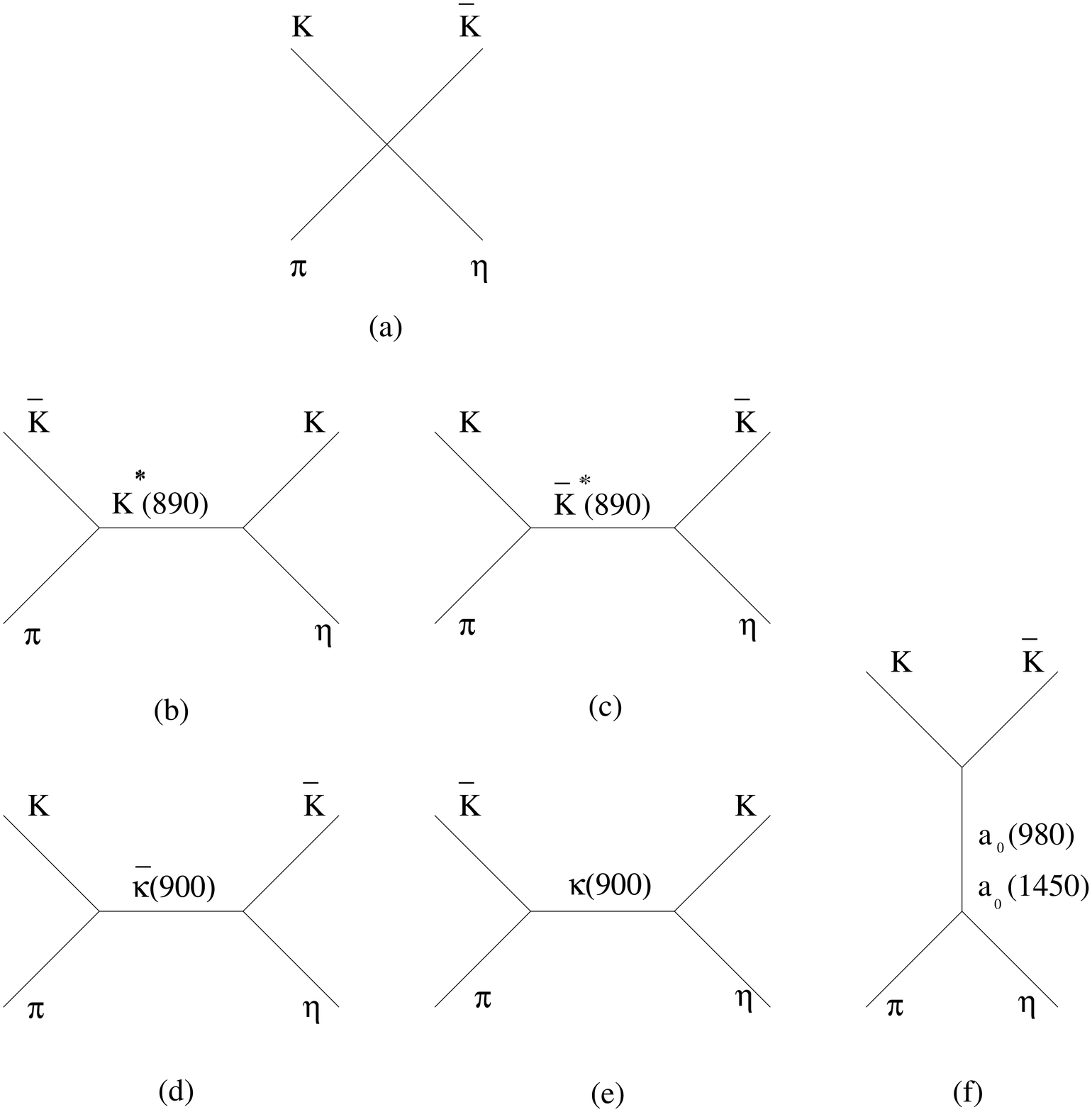, height=5in, angle=0}
\caption
{Feynman diagrams representing the contributions to the $\pi \eta
\rightarrow K \bar K$ scattering amplitude of (a) contact
terms, (b) and (c) vector mesons, (d) and (e) strange scalar $\kappa(900)$
mesons and (f) $a_0(980)$ and $a_0(1450)$ mesons.
}
\label{FD_pe2KK}
\end{figure}

Now we present an initial study of the $\pi \eta \rightarrow K \bar K$ and
$\pi \eta \rightarrow \pi {\eta}^{\prime}$ reactions in the energy range up
to about 1.6 GeV.  The Feynman diagrams needed to construct the $\pi \eta
\rightarrow K \bar K$ invariant amplitude in our model are shown in
Fig. \ref{FD_pe2KK}.  

Unlike the preceding $\pi \eta \rightarrow \pi \eta$ case, vector meson
exchanges and also contact terms arising from the first term of Eq. 
(\ref{Lag_psv}) are now allowed.  The contact contributions to the
amplitude for $A_{\pi \eta \rightarrow K \bar K} \left( s,t,u \right)$ for
${\pi}^+ \eta \rightarrow K^+ \bar{K^0}$ from the first term of
Eq. (\ref{Lag_psv}) (``current-algebra'' term) together with a piece from
the third term of Eq. (\ref{Lag_psv}) (due to addition of vector mesons in
a chiral symmetric way) are:
\begin{equation}
\frac{2s - t-u}{3F_{\pi}^2} \left( \frac {{\rm cos}\theta_p}{\sqrt 2} +
{\rm sin} \theta_p \right) \left[ 1 - \frac{ 3g_{\rho\pi\pi}^2 F_{\pi}^2}{4
m_{\rho}^2} \right].
\label{KKcontact}
\end{equation}
The second term in the square bracket (from the vectors) is about 1.6 times
the current algebra piece so it reverses the sign exactly as in the cases
of $\pi \pi$ and $\pi K$ scattering \cite{Sannino,Black1}.  The contact
amplitude arising from the pseudoscalar mass splittings in Eq. (\ref{PSSB}) is
\begin{equation}
{2\over {3 F_\pi^2} } 
\left( {\rm cos}\theta_p {{m_\pi^2 + m_K^2} \over \sqrt{2}} - {\rm
sin}\theta_p m_K^2 \right),
\end{equation}and turns out to be completely negligible.  The
$K^*(890)$ exchange contributions shown in (b) and (c) of Fig. {8} are
found to be 
\begin{equation}
- \left( {g_{\rho\pi\pi}\over 2} \right)^2
\left({ { {\rm cos}\theta_p} \over \sqrt{2}} + {\rm sin}\theta_p \right)
\left[{ { u-s +  { {(m_\pi^2 - m_K^2)(m_K^2-m_\eta^2)}\over { m_{K^*}^2}}} 
\over {m_{K^*}^2-t}}\right] + \left( t \leftrightarrow u \right).  
\end{equation}
The $\kappa (900)$ exchange contribution ((d) and (e) of
Fig. \ref{FD_pe2KK}) turns out to be negligible but arises from the amplitude piece
\begin{equation}
{ {\gamma_{\kappa K\pi} \gamma_{\kappa K\eta}} \over 4}
\frac{(t-m_\pi^2-m_K^2)(t-m_K^2-m_\eta^2)}{m_{\kappa}^2 -t} + \left( t
\leftrightarrow u \right).
\end{equation}

Finally, the s-channel contributions from the $a_0(980)$  and
$a_0(1450)$, shown in (f) of Fig. {8}, are being described by the
{\it{regularized}} amplitude
\begin{equation}
\sum_a \, { {\gamma_{a \pi\eta} \gamma_{aKK}} \over 4}
{{(s-2m_K^2)(s-m_\pi^2-m_\eta^2)}\over {m_{a}^2 - s -i m_{a}
\Gamma^{tot}_{a}}},
\label{KKschannel}
\end{equation}
where the sum is over $a=a_0(980)$ and $a_0(1450)$.  The entire amplitude
for $\pi \eta \rightarrow K \bar K$ is the sum of the terms above and is
invariant under $t \leftrightarrow u$ exchange, as expected from charge
conjugation invariance.  

\begin{figure}
\centering
\epsfig{file=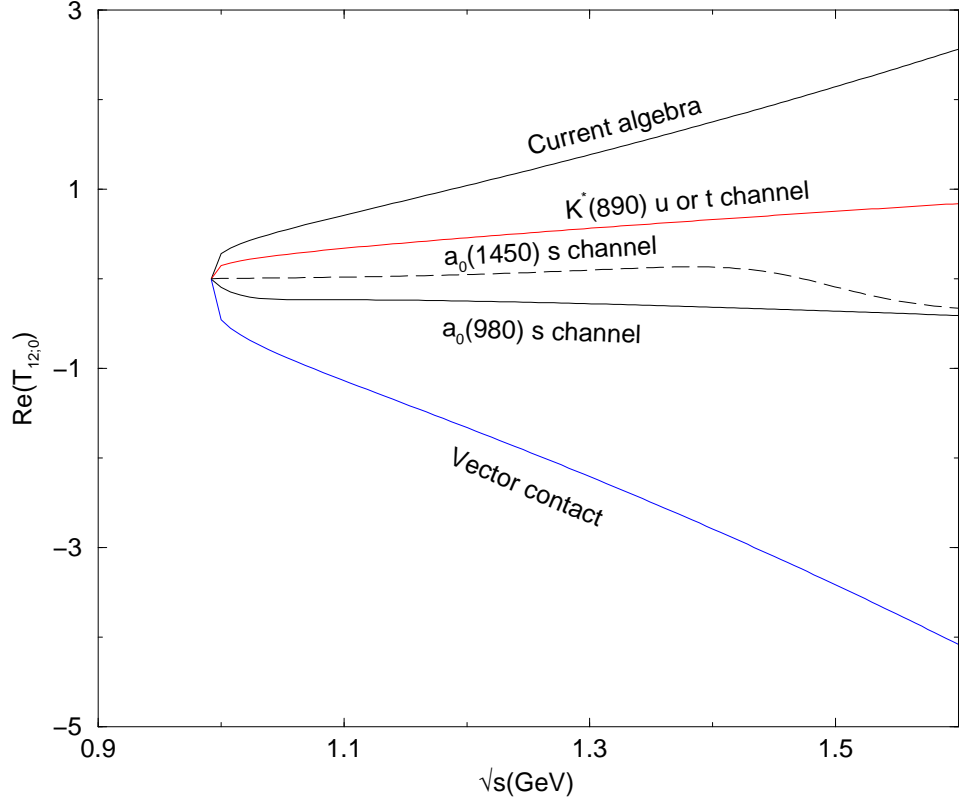, height=5in, angle=270}
\caption
{
Real parts of contributions to the $l=0$ partial wave amplitude for $\pi
\eta \rightarrow K \bar K$.
}
\label{pe2kk_indv}
\end{figure}

To proceed we have taken the projection of the amplitude into the $l=0$
partial wave using Eq. (\ref{projected}).  The real parts of the
non-negligible individual components are shown in Fig. \ref{pe2kk_indv}.
(Notice that the vertical scale has been greatly increased to accomodate
the rather large individual contributions here).   We see that
the contact terms are dominant although they partially cancel each other. 

\begin{figure}
\centering
\epsfig{file=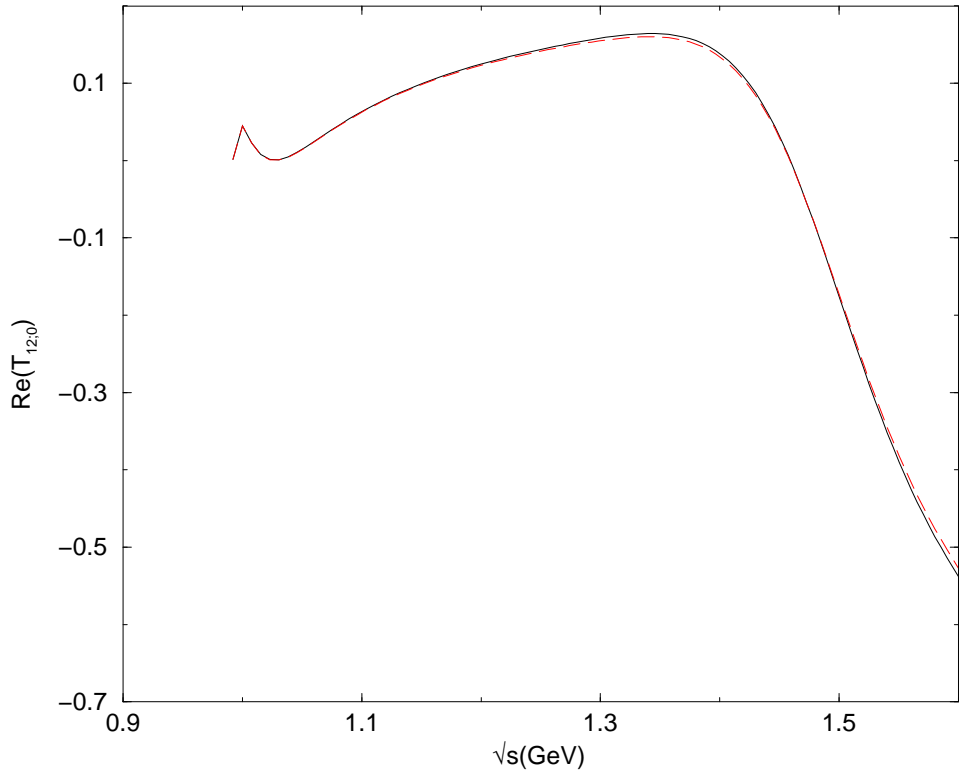, height=5in, angle=270}
\caption
{
Real part of the total $l=0$ partial wave amplitude for $\pi \eta
\rightarrow K \bar K$.
}
\label{pe2kk_Total}
\end{figure}

\begin{figure}
\centering
\epsfig{file=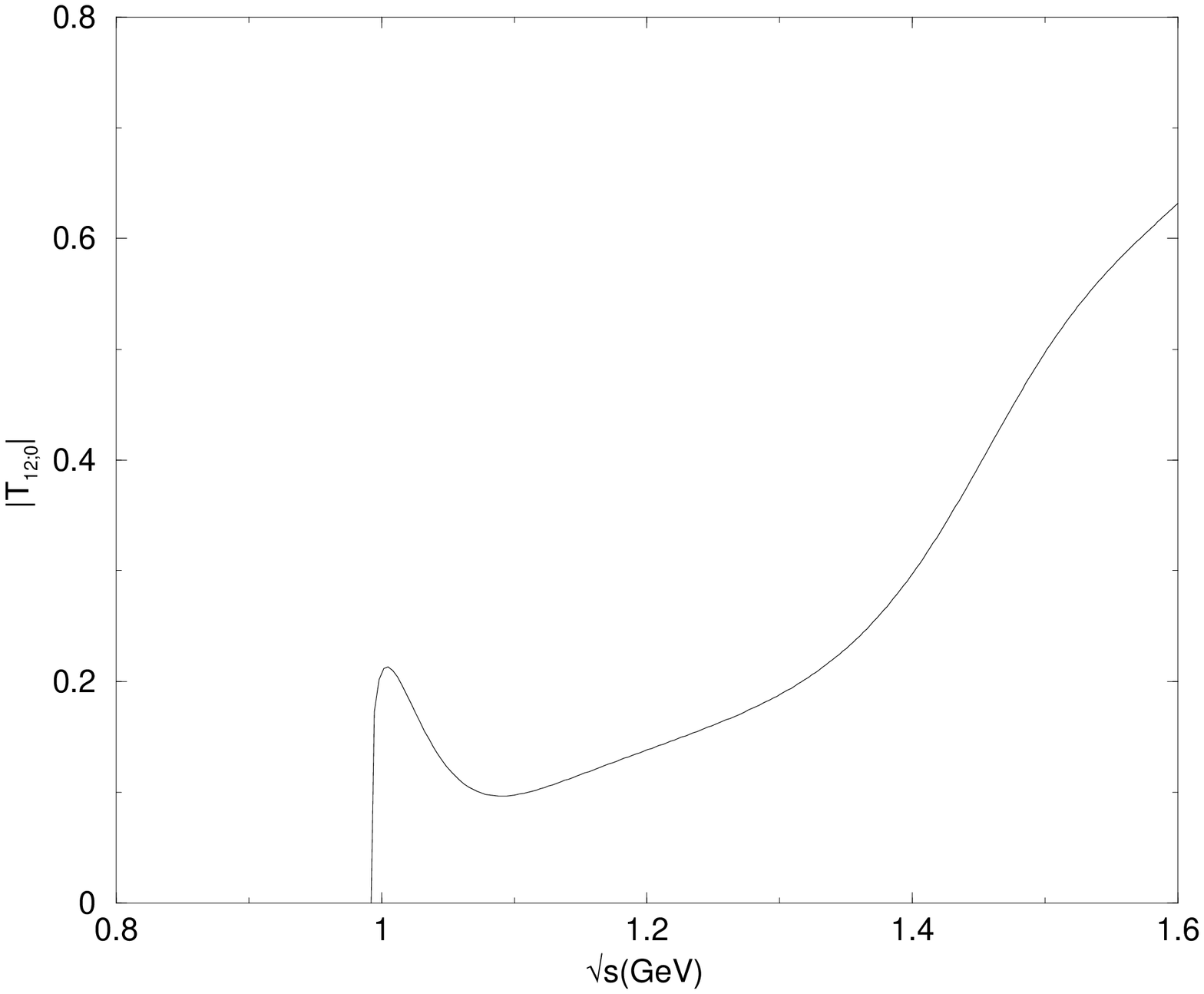, height=5in, angle=270}
\caption
{
Plot of the magnitude of the $l=0$ partial wave amplitude for $\pi \eta
\rightarrow K \bar K$.
}
\label{ModT12}
\end{figure}

The total real part of the $\pi \eta \rightarrow K \bar K$ partial wave is
shown in Fig. \ref{pe2kk_Total}.  Amusingly, all the large pieces in
Fig. \ref{pe2kk_indv} collaborate with each other to satisfy the unitarity
bound up to about 1.6 GeV.  To test this further we included the imaginary
parts of the partial wave amplitude, due to the regularization introduced
in Eq. (\ref{KKschannel}).  The plot is shown in Fig. \ref{ModT12}.  It is seen
that the stronger unitarity bound $\left| T_{12;0} \right| \leq \displaystyle{\frac{1}{2}}$ is violated only above 1.5 GeV.  

\begin{figure}
\centering
\epsfig{file=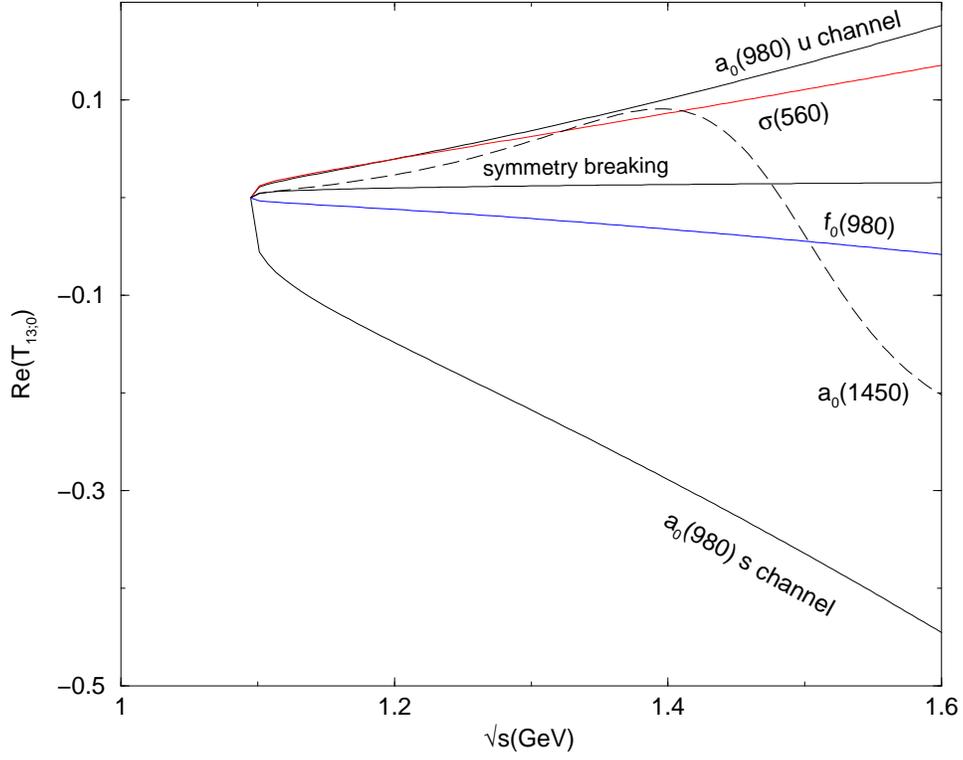, height=5in, angle=270}
\caption
{
Real parts of individual contributions to the $l=0$ partial wave amplitudes
for $\pi \eta \rightarrow \pi {\eta}^{\prime}$.
}
\label{pe2pep_indv}
\end{figure}

\begin{figure}
\centering
\epsfig{file=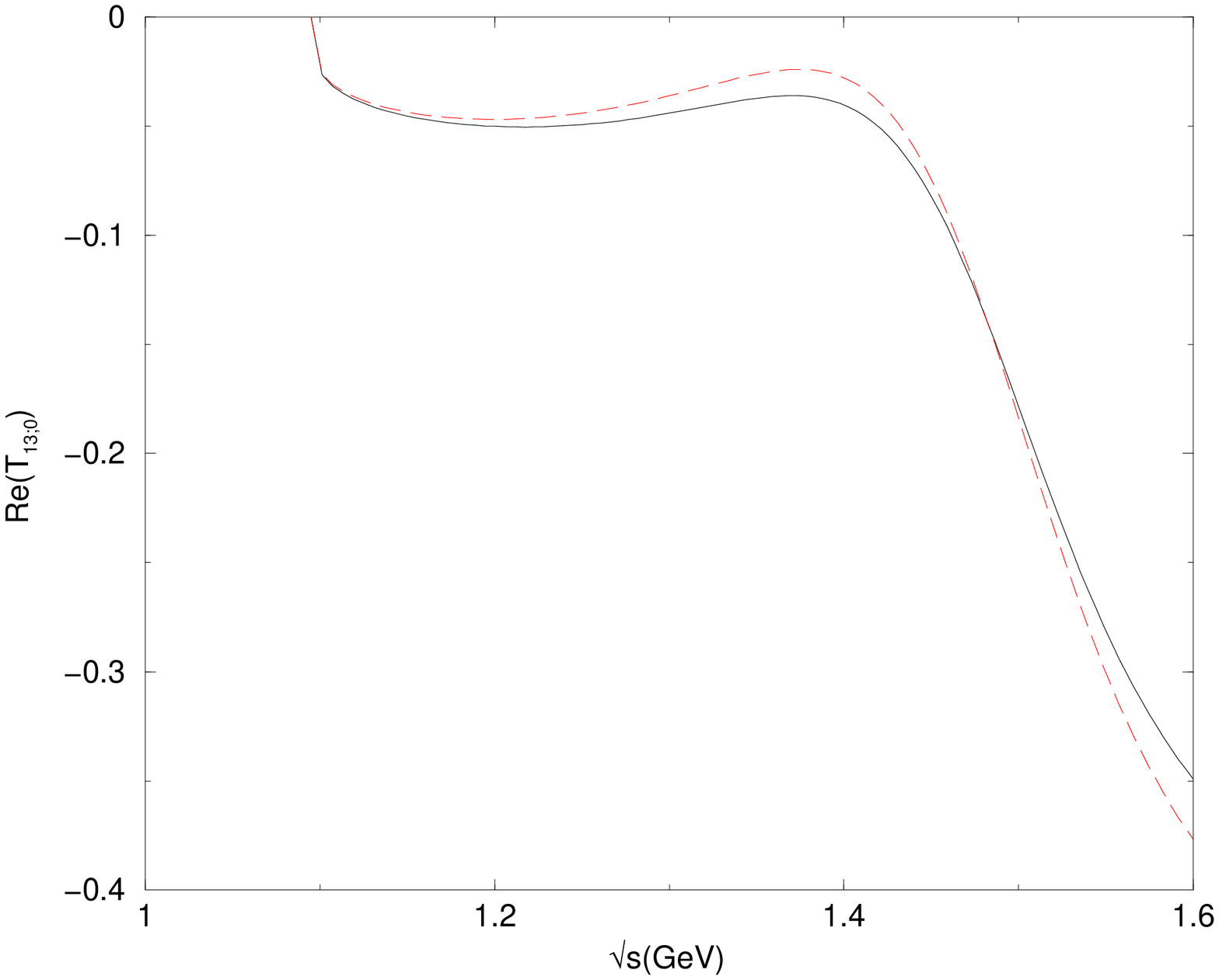, height=5in, angle=270}
\caption
{
Real part of the total $l=0$ partial wave amplitude for $\pi \eta
\rightarrow \pi {\eta}^{\prime}$.
}
\label{pe2pep_Total}
\end{figure}

\begin{figure}
\centering
\epsfig{file=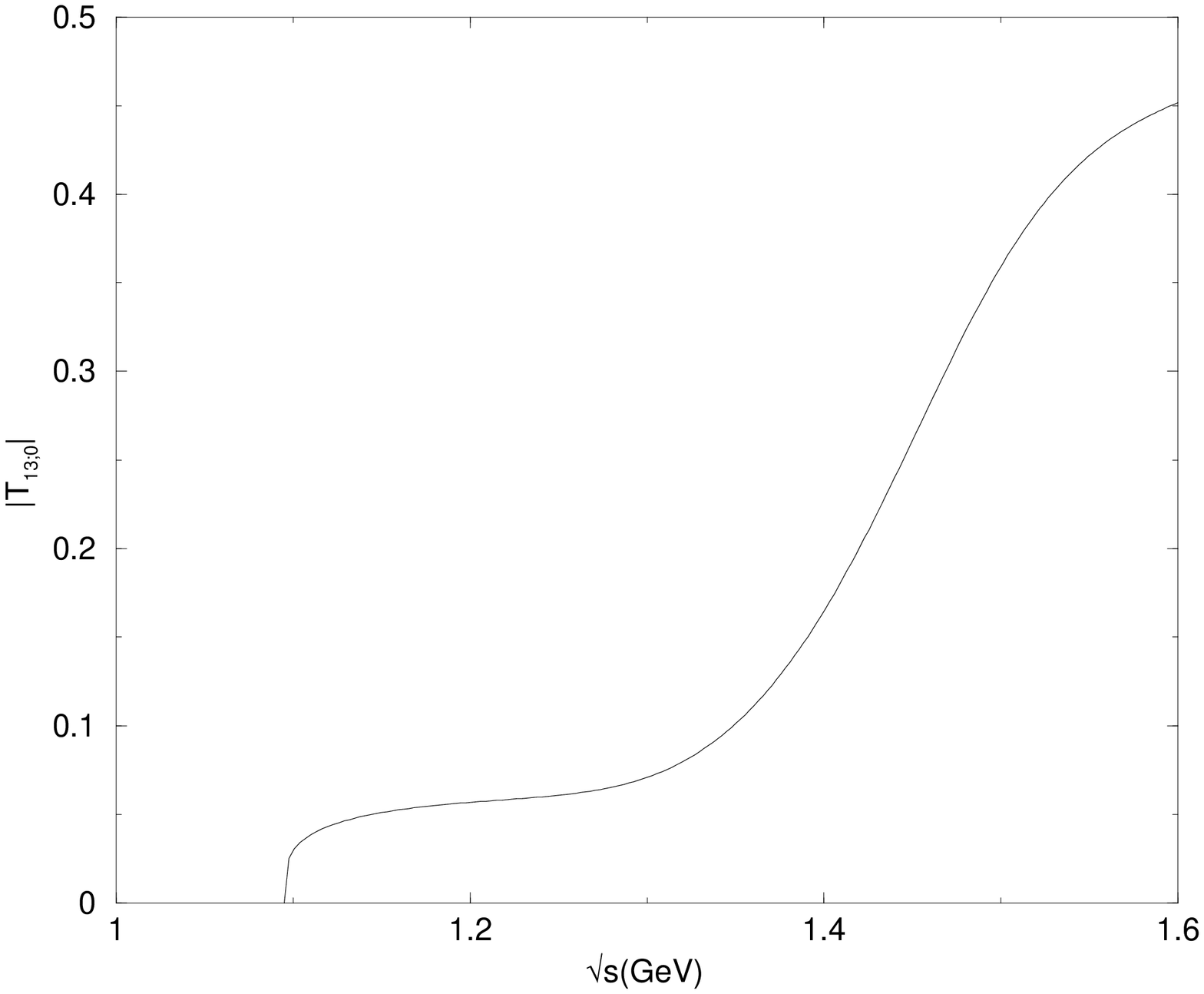, height=5in, angle=270}
\caption
{
Plot of the magnitude of the $l=0$ partial wave amplitude for $\pi \eta
\rightarrow \pi {\eta}^{\prime}$.
}
\label{ModT13}
\end{figure}

Finally let us consider the $\pi \eta \rightarrow \pi {\eta}^{\prime}$
amplitude.  The tree level Feynman diagrams are evident modifications of
those shown in Fig. {1} for the $\pi \eta \rightarrow \pi \eta$ case.  We
have the regularized (due to the $a_0(980)$ and $a_0(1450)$ poles) tree
amplitude 
\begin{eqnarray}
A_{\pi \eta \rightarrow \pi {\eta}^{\prime}} \left( s,t,u \right) &=&
\sum_{r=\sigma, f_0(980)} \frac{\gamma_{r\pi\pi}\gamma_{r\eta
{\eta}^{\prime}}}{\sqrt {2}} \frac{ \left( t - 2m_{\pi}^2 \right) \left( t
- m_{\eta}^2 - m_{{\eta}^{\prime}}^2 \right) }{m_r^2 - t } +  
\frac{m_{\pi}^2}{F_{\pi}^2}{\rm {sin}}2\theta_p  \nonumber \\
&+&\sum_{a} \frac{\gamma_{a\pi\eta}\gamma_{a\pi
{\eta}^{\prime}}}{4} \left[ \frac{ \left( s - m_{\pi}^2 - m_{\eta}^2 \right) \left( s
- m_{\pi}^2 - m_{{\eta}^{\prime}}^2 \right) }{m_a^2 - s - i m_a
\Gamma_a^{tot}} \right. \nonumber \\
 &+& \left. \frac{ \left( u - m_{\pi}^2 - m_{\eta}^2 \right) \left( u
- m_{\pi}^2 - m_{{\eta}^{\prime}}^2 \right) }{m_a^2 - u} \right],
\end{eqnarray}
where the a-summation goes over $a_0(980)$ and $a_0(1450)$.  The second term
is a small contact correction due to the pseudoscalar meson mass
splittings.  The individual contributions to the $l=0$ partial wave
projection of this amplitude are shown in Fig. \ref{pe2pep_indv} while the total $l=0$
projection is illustrated in Fig. \ref{pe2pep_Total}.  In this case, where the large
contact terms are absent, the unitarity bound $\left| {\rm {Re}} T_{13;0}
\right| \leq \frac{1}{2}$ is satisfied all the way up to 1.6 GeV.  This
also holds for the stronger bound, $\left| T_{13;0} \right| \leq \frac{1}{2}$, including the effects of the imaginary
part as shown in the plot of
Fig. \ref{ModT13}.  It is worthwhile to observe that the $a_0(1450)$ does
not dominate either of the off-diagonal $\pi \eta \rightarrow K \bar K$ or
$\pi \eta \rightarrow \pi {\eta}^{\prime}$ channels.  

\begin{figure}
\centering
\epsfig{file=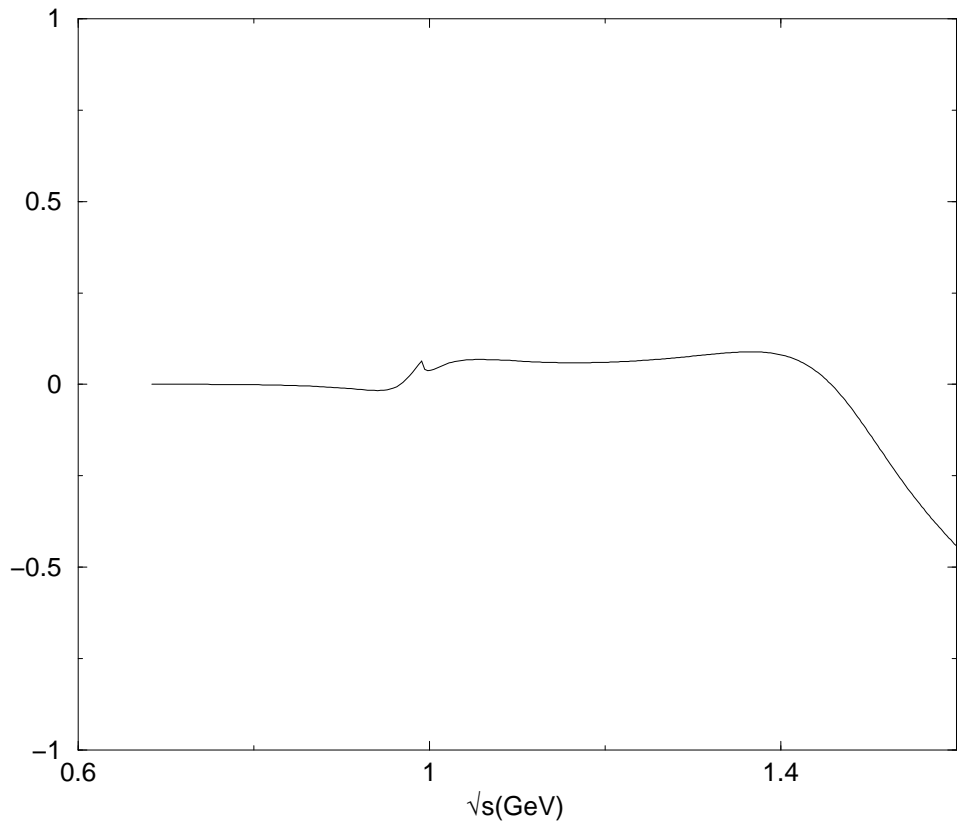, height=5in, angle=270}
\caption
{Plot of $\Delta$, the deviation from unitarity for $\pi \eta$ scattering
for $l=0$ channel.  $\Delta = 0$ for exact unitarity.
}
\label{U_test}
\end{figure}

Having looked at the individual $\pi \eta \rightarrow \pi \eta$, $\pi \eta
\rightarrow K \bar{K}$ and $\pi \eta \rightarrow \pi {\eta}^{\prime}$
channels it is now interesting to check the unitarity relation for these
l=0 S-matrix elements
\begin{equation}
\sum_{a} S_{1a;0}S_{1a;0}^* = 1,
\label{SSstar}
\end{equation}
on the assumption that the two-particle channels completely saturate the
s-wave $\pi \eta$ scattering at energies up to about 1.6 GeV.  Expressing
this formula in terms of the T-matrix elements suggests that we examine the
deviation,
\begin{equation}
\Delta = {\rm {Im}} \left( T_{11} \right) - {\left| T_{11} \right| }^2 -
{\left| T_{12} \right| }^2 - {\left| T_{13} \right| }^2
\end{equation}
which should vanish if Eq. (\ref{SSstar}) holds.  This is plotted in
Fig. \ref{U_test}.

It is seen that our simple model yields, up until about 1.4 GeV, relatively
small violations of unitarity for the s-wave amplitude.  This state of
affairs was obtained by projecting out the $l=0$ wave of an invariant
amplitude which is exactly crossing symmetric.  Thus the relative simplicity
of $\pi \eta$ scattering enables one to come closer to obtaining an amplitude which is both crossing symmetric and unitary.  

The larger violation of unitarity above 1.4 GeV can be seen to arise from
the violation of the unitarity bounds already noticed in the $\pi \eta
\rightarrow K \bar K$ amplitude (see Fig. \ref{ModT12}).  In turn this can be
traced to the relatively large contact and vector meson terms which
contribute to this particular off-diagonal process.  This is in contrast to the
$\pi \eta \rightarrow \pi \eta$ and $\pi \eta \rightarrow \pi
{\eta}^{\prime}$ amplitudes for which the potentially large current algebra
and vector meson terms were seen to vanish.  Of course the large current
algebra and vector meson pieces were very important in previous discussions
of the $\pi \pi$ and $\pi K$ scatterings.  In this sense, since they only
enter through the ``back door'' of an off-diagonal channel for $\pi \eta$
scattering, the $\pi \eta$ scattering is effectively simpler to treat
in our model.  

Here, as mentioned above, we have just made an initial exploration of the
coupled-channel $\pi \eta$ scattering problem.  A fuller treatment would include the exchanges of spin
2 and other higher mass resonances and also scattering processes having
$K \bar K$ and $\pi \eta^{\prime}$ initial states.  For example important
contributions in the $K \bar K$ case may be expected from the $K^*(1410)$.

\section{Discussion}

We studied $\pi\eta$ scattering in a model where the starting amplitude
was computed at tree level from a phenomenological chiral Lagrangian
containing exchanges of resonances having masses within the energy range
of interest.  Divergences at the direct channel poles were regularized in
a conventional manner.   Previously this method was applied to $\pi\pi$
scattering and $\pi K$ scattering; consistency required the existence of
$\sigma(560)$ and $\kappa(900)$ scalar resonances.  From that work  and
also from a related study of $\eta'\rightarrow \eta\pi\pi$ decay, all the 
light scalar-pseudoscalar-pseudoscalar coupling constants were determined.
These were used without change in the present work, which essentially
involves no new parameters.

We first examined the roughly elastic region (up to about 1.2 GeV) which 
is dominated by the $a_0(980)$ resonance.   It is noteworthy that neither
vector mesons nor large ``current algebra'' contact terms can contribute
in this region, unlike the $\pi\pi$ and $\pi K$ cases.   The non-trivial
contributions all arise from light scalar meson exchanges.   Thus $\pi
\eta$ scattering seems an excellent channel for learning more about these 
resonances which are of great current interest.   Our model in section II  was
exactly crossing symmetric, but not exactly unitary.  To see why that
model which features an $a_0(980)$ with mass and width consistent with the
experimental values is not automatically unitary we note the following two
points:  First, the amplitude is more general than a pure Breit-Wigner pole for 
the $a_0(980)$ s-channel since $\sigma$ and $f_0$ t-channel exchanges and
$a_0(980)$ u-channel exchange also exist.   Furthermore chiral symmetry
dictates a non-trivial momentum dependence of the coupling factors.

To further investigate this question we constructed (in Section III) a
related partial wave amplitude which we made unitary without regard for
crossing symmetry.  A comparison of the unitary and the crossing symmetric
amplitudes showed that they were in fact close; the difference between
them gives an estimate of the uncertainty of our approach.  

It is encouraging to us that treating the $a_0(980)$ in $\pi \eta$
scattering by the same method as used in earlier discussions of $\pi \pi$
scattering (in which a $\sigma (560)$ and the $f_0(980)$ appeared) and $\pi
K$ scattering (in which a $\kappa (900)$ was needed) seems to be
reasonable.  Now, all the members of a possible low-lying scalar nonet have
been studied through their appearance in meson-meson scattering.   Of
course, many questions remain.  One concerns the ``family'' structure of
such a possible nonet; we have discussed some speculations on this aspect
elsewhere \cite{Black2,mechanism}.  From the present viewpoint, the most
important aspect concerns improving our understanding of the meson-meson
chiral scattering amplitude.  Clearly, one way to proceed is to examine the
model at higher energies.  

We thus made a preliminary exploration in section IV of the nearby
inelastic region (roughly $1-1.5$ GeV).  This range features the
$a_0(1450)$ scalar resonance.  It turns out that the earlier terms produce
essentially zero background interference at the position of the
$a_0(1450)$.  Thus, it is natural to add this particle into the picture
directly, yielding a crossing symmetric amplitude.  This amplitude was seen
to satisfy the unitarity bounds when the effects of inelasticity were
incorporated in the regularization of the $a_0(1450)$ pole.  The incomplete
experimental data on the $a_0(1450)$  branching ratios were interpreted
with the aid of a simple model.  Similar preliminary discussions were given
for the $\pi \eta \rightarrow K \bar K$  and $\pi \eta \rightarrow \pi
{\eta}^{\prime}$ off-diagonal processes.  An additional complication showed
up in the $\pi \eta \rightarrow K \bar K$ case.  Here the vector meson
$K^*$ exchange and a large current algebra contact term both contribute as
in the $\pi \pi$ and $\pi K$ scatterings.  Nevertheless it was found that
the exactly crossing symmetric amplitudes for $\pi \eta \rightarrow \pi
\eta$, $K \bar K$ and $\pi {\eta}^{\prime}$ satisfied the unitarity
relation amongst themselves to a reasonable accuracy until about 1.4 GeV.  

There are many interesting directions for future work.  Clearly a full
3-channel analysis and investigation of various alternative unitarization
schemes is suggested.  Inclusion of the many resonances in the $1-1.5$ GeV
range which can be exchanged is also desirable.  Especially interesting are
the effects of the isoscalar scalars in this energy regime.  Together with
parallel expanded treatments of the $\pi \pi$ and $\pi K$ cases we may hope
to learn about a second possible scalar nonet containing the $a_0(1450)$
and $K_0^*(1430)$.  The knowledge of the scalar coupling constants obtained
from refined analyses can be useful in the treatment of other physical
processes.  The recently measured \cite{VEPP} radiative decay $\phi(1020)
\rightarrow a_0(980)+\gamma$ is especially important.

\acknowledgments
We would like to thank Francesco Sannino for helpful discussions.  
The work has been supported in part by the US DOE under contract
DE-FG-02-85ER 40231.

\appendix

\section{Chiral Lagrangian}
First we write the chiral Lagrangian of pseudoscalars and vectors.  This
makes use of the $3 \times 3$ matrix $U=e^{2 i\frac{\phi}{F_\pi}}$ wherein
$\phi$ represents the usual $3 \times 3$ matrix of pseudoscalar fields.
Defining the square root $\xi$ by $U = \xi \xi$ we consider the
combinations
\begin{eqnarray}
p_\mu &=& \frac{i}{2}
\left( 
  \xi \partial_\mu \xi^{\dag} - \xi^{\dag} \partial_\mu \xi
\right) \ ,
\nonumber\\
v_\mu &=& \frac{i}{2}
\left( 
  \xi \partial_\mu \xi^{\dag} + \xi^{\dag} \partial_\mu \xi
\right) \ .
\end{eqnarray}
Under a chiral transformation $U \rightarrow U_{\rm L} U U_{\rm R}^{\dag}$,
$\xi$ transforms as:
\begin{equation}
\xi \rightarrow 
U_{\rm L} \, \xi \, K^{\dag}(\phi,U_{\rm L},U_{\rm R}) = 
K(\phi,U_{\rm L},U_{\rm R}) \, \xi \, U_{\rm R}^{\dag}, \ 
\end{equation}
which defines $K(\phi,U_{\rm L},U_{\rm R})$.  A vector meson nonet
$\rho_\mu$ transforms as a gauge field
\begin{equation}
\rho_\mu \rightarrow K \rho_\mu K^{\dag} + 
\frac{i}{\widetilde{g}} K \partial_\mu K^{\dag}.
\end{equation}
Similarly
\begin{equation}
v_\mu \rightarrow K v_\mu K^{\dag} + i K \partial_\mu K^{\dag},
\end{equation}
while 
\begin{equation}
p_\mu \rightarrow K p_\mu K^{\dag}. 
\end{equation}
The chiral invariant (neglecting quark mass induced terms) Langrangian of
pseudoscalars and vectors may then be written as 
\begin{equation}
{\cal L}_1 = - \frac{F_\pi^2}{8} {\rm {Tr}}\left( \partial_\mu U
{\partial_\mu U}^{\dagger} \right)-\frac{1}{4} \mbox{Tr} 
\left[ F_{\mu\nu}(\rho) F_{\mu\nu}(\rho)\right] -\frac{1}{2} m_{\rho}^2 \mbox{Tr} 
 \left(\rho_\mu - \frac {v_\mu}{\widetilde{g}} \right)^2 -
\frac{m_0^2 F_\pi^2}{96} {\left( {\rm {ln}} \frac{ {\rm {det}}
U}{{\rm{det}}{U}^{\dagger}} \right) }^2, 
\label{Lag_psv}
\end{equation}
where
$F_{\mu\nu} = \partial_\mu \rho_\nu - \partial_\nu \rho_\mu - i \widetilde{g} 
[ \rho_\mu , \rho_\nu ]$ is the vector meson {\it gauge field strength}.
$F_{\pi} = 0.131$ GeV is the pion decay constant, $\tilde{g}$ is the vector
meson gauge coupling constant and $m_0$  is a mass for the unmixed
${\eta}^{\prime}$ field.  $\tilde{g}$ is related to the $\rho$ meson width
by
\begin{equation}
\Gamma \left( \rho \rightarrow \pi \pi \right) =
\frac{g_{\rho\pi\pi}^2{q_{\pi}^3}}{12 \pi m_{\rho}^2} \quad , \quad
g_{\rho\pi\pi} = \frac{m_\rho^2}{\tilde{g} F_\pi^2} \approx 8.56
\end{equation}
and $m_\rho= 0.77$ GeV.  $q_\pi$ is the pion momentum in the rest frame of
the decaying $\rho$ meson.  The last term in Eq. (\ref{Lag_psv}) reflects
the $U\left( 1 \right) _A$ anomaly in QCD and allows the ${\eta}^{\prime}$
particle to have a suitably large mass.  We take $m_\pi = 0.137$ GeV, $m_K
= 0.496$ GeV, $m_\eta = 0.547$ GeV, $m_{{\eta}^{\prime}} = 0.958$ GeV and
$m_{K^*} = 0.890$ GeV.

Next consider the chiral invariant Lagrangian involving the scalar nonet
field $N$.  We note \cite{CCWZ} that $N$ may be taken to transform as 
\begin{equation}
N \rightarrow K N K^{\dagger}.
\end{equation}
Then it is clear that  
\begin{eqnarray}
{\cal L} &=& - \frac{1}{2}{\rm Tr}\left( {\cal D}_\mu N {\cal D}_\mu N \right)
- a {\rm Tr} \left( N N \right) - c{\rm Tr} \left( N \right) {\rm Tr} \left( N \right) + F_\pi^2 \left[A\epsilon^{abc}\epsilon_{def}N_a^d {\left( p_\mu \right)}_b^e{\left( p_\mu \right)}_c^f \right.  \nonumber \\ 
&+& \left. B{\rm Tr}\left( N \right) {\rm Tr}\left( p_\mu p_\mu \right) + C {\rm Tr}  \left( N p_\mu \right) {\rm Tr}\left( p_\mu \right) + D {\rm Tr} \left( N \right) {\rm Tr} \left( p_\mu \right) {\rm Tr} \left( p_\mu \right) \right],
\label{Lag_scalars}
\end{eqnarray}
where ${\cal D}=\partial_\mu - i v_\mu$, is chirally invariant.  The
coupling constants describing scalar $\rightarrow$ pseudoscalar $+$
pseudoscalar are given in terms of the coefficients $A$, $B$, $C$ and $D$;
these were determined from $\pi \pi$ scattering, $\pi K$ scattering and
${\eta}^{\prime} \rightarrow \eta \pi \pi$ decay in \cite{Black2,Fariborz}.
As an explicit example of a coupling constant describing a trilinear
interaction among isomultiplets extracted from Eq. (\ref{Lag_scalars})
consider
\begin{equation}
-{\cal L}_{N\phi\phi} = \frac{\gamma_{\kappa K \pi}}{\sqrt 2} \left(
\partial_\mu {\bar K} \mbox{\boldmath ${\tau}$} \cdot \partial_\mu {\mbox{\boldmath ${\pi}$}}
\kappa + h.c. \right) + ...,
\end{equation}
which yields the identification $\gamma_{\kappa K \pi} = -2A$.  The other
terms in this isotopic spin decomposition are also given in \cite{Black2}.

Symmetry breaking terms must still be added.  These involve the ``spurion''
matrix
\begin{equation}
{\cal M} = {\rm diag} \left( 1,1,x \right)
\end{equation}
where $x \approx 20.5$ \cite{Harada-Schechter} is the strange to nonstrange quark mass ratio.  Pseudoscalar mass
terms are propotional to 
\begin{equation}
{\rm Tr} \left( U {\cal M}^{\dagger} \right ) + h.c.,
\label{PSSB}
\end{equation}
vector mass splitting terms are contained in a term
\begin{equation}
{\rm Tr} \left[ {\xi}^{\dagger} {\cal M} \xi^\dagger {\left( \tilde{g}
\rho_\mu - v_\mu \right)}^2 \right] + h.c.
\end{equation}
and finally scalar meson mass splittings are contained in:
\begin{equation}
-\frac{b}{2} {\rm Tr} \left( NN\xi^\dagger {\cal M}\xi^\dagger \right) +
\frac{d}{2} {\rm Tr}\left( N \right){\rm Tr} \left( N  \xi^\dagger {\cal M}\xi^\dagger \right) +h.c.
\end{equation}

A detailed discussion of the scalar meson mass terms and the determination
of the parameters $a$, $b$, $c$ and $d$ was given in \cite{Black2}.

The main effects of these mass splitting terms are to give each particle
its correct experimental mass and to accomodate mixing between particles of
the same spin-parity and isospin.  Our conventions for the $\eta -
{\eta}^\prime$ and $\sigma-\sigma^\prime$ mixing are:

\begin{equation}
\left( 
\begin{array}{c} 
        \eta\\ 
        \eta' 
\end{array} 
\right) =
\left( 
\begin{array}{c c} 
{\rm cos} \theta_p  & -{\rm sin} \theta_p \\
{\rm sin} \theta_p  &  {\rm cos} \theta_p 
\end{array} 
\right)
\left( 
\begin{array}{c} 
 (\phi^1_1+\phi^2_2)/ \sqrt{2} \\ \phi^3_3 
\end{array} 
\right)
\label{eta-etap}
\end{equation}
and
\begin{equation}
\left( \begin{array}{c} \sigma\\ f_0 \end{array} \right) = \left(
\begin{array}{c c} {\rm cos} \theta_s & -{\rm sin} \theta_s \\ {\rm sin}
\theta_s & {\rm cos} \theta_s \end{array} \right) \left( \begin{array}{c}
N_3^3 \\ \frac {N_1^1 + N_2^2}{\sqrt 2} \end{array} \right).
\label{mixing-convention}
\end{equation}

The asymmetry in these two definitions reflects the prejudice that in the
ideal mixing limit (zero mixing angle) the heaviest pseudoscalar
$\eta^\prime$ is identified as $\phi_3^3$ while the lightest scalar
$\sigma$ is identified as $N^3_3$.  We choose the conventional value
$\theta_p = 37^{o}$ and a value $\theta_s = -20.3^{o}$ as
discussed in \cite{Black2}.  This corresponds to the scalar meson masses
(needed to explain $\pi\pi$ and $\pi K$ scattering in our model):
\begin{eqnarray}
m_\sigma = 550 \, {\rm MeV} \,\, &,& \, \, m_{f_0} = 980\, {\rm MeV} \nonumber \\
m_{a_0} = 983.5 \, {\rm MeV} \, \, &,& \, \, m_\kappa = 897 \, {\rm MeV}.
\end{eqnarray}

We take the heavier scalar isovector to have a mass given by \cite{PDG},
namely $m[a_0(1450)] = 1.474$ GeV.  The coupling constants for the light
scalars to two pseudoscalars relevant for the present paper are (in units
of ${\rm{GeV}}^{-1}$)
\cite{Black2,Fariborz}:
\begin{eqnarray}
\gamma_{\sigma\pi\pi} = 7.27 \, \, &,& \, \, \gamma_{\sigma\eta\eta} = 4.11
\, \, , \, \, \gamma_{\sigma\eta{\eta}^{\prime}} = 2.65 \nonumber \\
\gamma_{f_0\pi\pi} = 1.47 \, \, &,& \, \, \gamma_{f_0\eta\eta} = 1.72
\, \, , \, \, \gamma_{f_0\eta{\eta}^{\prime}} = -9.01 \nonumber \\
\gamma_{a_0\pi\eta} = -6.80 \, \, &,& \, \, \gamma_{a_0 K \bar K} = -5.02
\, \, , \, \, \gamma_{a_0\pi{\eta}^{\prime}} = -7.80 \nonumber \\
\gamma_{\kappa K \pi} && = -5.02 \, \, , \gamma_{\kappa K \eta} = -0.94
\end{eqnarray}

\section{Partial wave projection of $\pi \eta \rightarrow \pi \eta$
elastic amplitude}

The s-wave projections of the invariant amplitudes represented in
Fig. \ref{FD_pe2pe} are calculated from the $l=0$ and $a=b=1$ case of
Eq. (\ref{projected}).  The direct channel s-wave amplitude is:
\begin{equation}
T_{11;0}^{s-channel} = \rho (s) \frac {{\gamma_{a_0\pi\eta}}^2}{2} \frac {{\left( s - m_\pi^2 -
m_\eta^2 \right) }^2}{m_{a_0}^2 - s -im_{a_0}G_{a_0}^{\prime}\theta \left[s
- {\left(m_\eta + m_\pi \right)}^2 \right]},
\end{equation}  
where $\rho (s) \equiv \rho_1 (s)$ in the notation of
Eq. (\ref{kinematical}) for channel 1 ($\pi \eta \rightarrow \pi\eta$).
The projection of the t-channel amplitude of Fig. \ref{FD_pe2pe}(a) is:
\begin{equation}
T_{11;0}^{t-channel, r} = \rho(s) q^2 \frac{\gamma_{r\pi\pi}
\gamma_{r\eta \eta}}{\sqrt 2} \left[  2\alpha - 4\gamma
+ \left(  \alpha \gamma - \beta - 2 \gamma^2
\right) {\rm {ln}} \left| \frac{\gamma - 1} {\gamma + 1} \right| \right],
\end{equation}
where $q^2$ is the center of mass momentum and 
\begin{equation}
\alpha \equiv 2 \frac{m_\pi^2 + m_\eta^2 + 2q^2}{q^2} \, , \, \beta \equiv
2 \frac {\left(m_\pi^2 + q^2 \right) \left(m_\eta^2 + q^2 \right)}{q^4} \, , \, \gamma \equiv
\frac {m_r^2 + 2q^2}{2q^2}.
\end{equation}  
Here there is one term with $r=\sigma$ and another with $r=f_0(980)$.
Finally, the $l=0$ projection of the $a_0(980)$ u-channel exchange amplitude is:
\begin{equation}
T_{11;0}^{u-channel} = \rho(s) q^2 \frac{{\gamma_{a\pi\eta}}^2}{2}
\left[ 2B + 4C + C^2 {\rm {ln}} \left| \frac{B+1}{B-1} \right| \right],
\end{equation}
where
\begin{equation}
B \equiv \frac{1}{2q^2} \left[ m_a^2 - m_\eta^2 - m_\pi^2 + 2 \sqrt{m_\pi^2
+ q^2}{\sqrt {m_\eta^2 + q^2}}\right] \equiv \frac{{\sqrt {m_\pi^2 +
q^2}}{\sqrt {m_\eta^2 + q^2}}}{q^2} - C.
\end{equation}

\end{document}